\documentclass[1p, hidelinks]{elsarticle} 

\usepackage{lineno,hyperref}

\usepackage{xcolor}

\usepackage{geometry}
\usepackage{graphicx} 
\usepackage{amsmath}
\usepackage{amsthm}
\usepackage{mathtools}
\usepackage{amssymb}
\usepackage{caption}
\usepackage[toc,page]{appendix}
\usepackage{booktabs}
\usepackage{multirow}
\usepackage{bm}

\modulolinenumbers[5]

\journal{Structural Safety}

\makeatletter
\def\ps@pprintTitle{%
 \let\@oddhead\@empty
 \let\@evenhead\@empty
 \def\@oddfoot{}%
 \let\@evenfoot\@oddfoot}
\makeatother
\newcommand\blfootnote[1]{%
  \begingroup
  \renewcommand\thefootnote{}\footnote{#1}%
  \addtocounter{footnote}{-1}%
  \endgroup
}

\newcommand{\vor}[1]{\textrm{Vor}(#1)}

\newcommand{\norm}[1]{\left\lVert#1\right\rVert}

\newtheorem{theorem}{Theorem}[section]
\newtheorem{lemma}[theorem]{Lemma}
\newtheorem{corollary}[theorem]{Corollary}
\newtheorem{proposition}[theorem]{Proposition}
\newtheorem{alg}[theorem]{Algorithm}
\newtheorem{definition}[theorem]{Definition}

\begin{document}

\begin{frontmatter}

\title{Environmental contours as Voronoi cells}

\author[dnvgl]{Andreas Hafver}
\author[dnvgl,uio]{Christian Agrell}
\author[dnvgl,uio]{Erik Vanem}

\address[dnvgl]{DNV GL - Group Technology and Research, Norway}
\address[uio]{Department of Mathematics, University of Oslo, Norway}

\blfootnote{\textbf{pre-print:} This is a pre-print version of this article}

\begin{abstract}

    Environmental contours are widely used as basis for design of structures exposed to environmental loads. The basic idea of the method is to decouple the environmental description from the structural response. 
    This is done by establishing an envelope of environmental conditions, such that any structure tolerating loads on this envelope will have a failure probability smaller than a prescribed value. 
    
    Specifically, given an $n$-dimensional random variable $\mathbf{X}$ and a target probability of failure $p_{e}$, 
    an \textit{environmental contour} is the boundary of a set $\mathcal{B} \subset \mathbb{R}^{n}$ with the following property:
    For any \textit{failure set} $\mathcal{F} \subset \mathbb{R}^{n}$, if $\mathcal{F}$ does not intersect the interior of $\mathcal{B}$, 
    then the probability of failure, $P(\mathbf{X} \in \mathcal{F})$, is bounded above by $p_{e}$.
    As is common for many real-world applications, we work under the assumption that failure sets are convex.
    
    In this paper, we show that such environmental contours may be regarded as boundaries of Voronoi cells. This geometric interpretation leads to new theoretical insights and suggests a simple novel construction algorithm that guarantees the desired probabilistic properties. 
    The method is illustrated with examples in two and three dimensions, but the results extend to environmental contours in arbitrary dimensions.
    Inspired by the Voronoi-Delaunay duality in the numerical discrete scenario,  we are also able to derive an analytical representation 
    where the environmental contour is considered as a differentiable manifold,
    and a criterion for its existence is established. 
    
    
\end{abstract}


\end{frontmatter}


\section{Introduction and background}

\subsection{A brief review of environmental contours}

The use of environmental contours is a well-established practice in design of marine structures, 
and helps the designer identify design sea states corresponding to extreme environmental loads associated with a certain return period. 
The concept of environmental contours is an efficient method for estimating multivariate extreme conditions, 
and it is an alternative to full long-term response analysis in situations where this is not feasible. 
The environmental contour method is also recommended in standards and recommended practices such as \cite{DNVGLRP-C205, NORSOK_N003_17}. 

The concept of environmental contours was first introduced by \cite{Haver80, Haver87} as 
a means to study the joint distribution of significant wave height and wave period of ocean 
waves. These early environmental contours were based on constant densities, but the concept of environmental 
contours was developed further by \cite{WitUCBH93} by using the Inverse First Order Reliability Method (IFORM) 
and considering exceedance probabilities in the transformed standard normal space \cite{HW:ENvContLin09}. 
The IFORM method avoids unnecessary conservatism in the equi-density contours \citep{Leira:StocProcCont08}, and has since then become the most applied contour method. Several applications of the environmental contour method in marine engineering and design are reported in the literature \citep{NdLY:EstExtrRespEnvCont98, WJK:RelFloatStructLoadFactDesign99, BM:ApplyContHullLoads01, SM:DesignLoadWindTurbEnvCont06, BHL:OMAE2007-29417, BH_EnvContSum09, bhoe:CombContHsT10, OMAE2011-49886, HaverWAveWorkshop13, MGM:ApplContTwoBodFloatEngConv13}. A comparison study presented in \cite{OMAE2015-41680} investigated the influence of the choice of contour method on some vessel responses. 

Environmental contours continues to be an active area of research, and several modified approaches have been suggested in recent years, e.g., a dynamical IFORM method \citep{LW:DynamicIFORM16}, a modified approach to account for non-monotonic behaviour of the responses \citep{LGM:ModEnvCont16}, an approach including pre-processing and principal component analysis prior to estimating IFORM contours \citep{E-GSDN:PCAContours16}, contours for sub-populations such as directional sectors or seasonality \citep{VanemSeasonalContour17, HusebyVB:ESREL2019}, contours for a combination of circular and linear variables \cite{HK:EnvContCircLin17}, contours for copula-based joint distributions \citep{S-GH-ZM-I:EnvContNATAF13, M-IH-Z:EnvContCop15} and contours based on a direct IFORM approach \cite{DH:NewEnvCont:OMAE2019-95993}. Contours for buffered failure probabilities were proposed in \cite{DahlHuseby:ESREL18} and contours based on a particular version of the inverse second order reliability method (ISORM) were derived in \cite{CL:ISORMContours18}. Recently, the initial equi-density method was revisited in \cite{HOWT:HighDensityContour2017}. The uncertainties associated with environmental contours due to uncertainties in the underlying joint distribution model and due to sampling variability are investigated in \cite{M-IHZ:UncertEnvCont17} and \cite{VanemGB-G:ContourUncert18}, respectively, and weighted environmental contours based on combining data from different datasets were explored in \cite{Vanem:CombinedContours19}. Reviews of various contour methods are presented in e.g. \cite{MNCCE-GM:AltApproachEnvCont18, ECSADESjointPaper19}. 

An alternative approach to constructing environmental contours that avoids the transformation into standard normal space, but rather defines exceedance probabilities in the original parameter space, was proposed in \cite{Vanem:EnvCont12, Vanem:EnvCont14}. This is based on Monte Carlo simulations from the joint distribution of environmental parameters, and initial inaccuracies due to insufficient number of Monte Carlo samples were overcome by a scheme for tail sampling as outlined in \cite{HusebyVN:ESREL2014}. It is argued that the contours obtained in this way have more well defined probabilistic properties, and an evaluation of the properties of the IFORM-based environmental contours is presented in \cite{HusebyVE:ESREL2017}.  However, in some situations it is found that the direct sampling contours may contain irregularities in the form of small loops, as discussed in \cite{Vanem:EnvCont14}. One reason for this is related to the Monte Carlo variance and the fact that the contours are estimated based on a finite sample from the joint distribution, and the issue may be resolved by increasing the number of Monte Carlo samples. However, the reason may also be genuine features of the underlying joint distribution, i.e. that the joint distribution does not admit a proper convex environmental contour.
A comparison study on the IFORM and the Monte Carlo-based approach to environmental contours was presented in \cite{VanemB-G:JOMAE15}, which demonstrated that in certain cases, notable different contours are obtained. The comparison study was extended to consider various simple structural problems in \cite{VanemContourCompStruc17} and to compare contour-based methods to response-based methods in \cite{ECSADESShipCase19}. 

Even though many structural problems depend on more than two environmental variables, 
most applications of environmental contours are restricted to two-dimensional contours. 
For example, in the multivariate problem addressed in \cite{NFPQ-R:JointDistMultVarIFORMCont07}, 
environmental contours were only calculated for pairs of variables. However, some examples of three-dimensional contours based on the IFORM approach, are shown in \cite{dLN:EnvContEarthquake00, SM:ModWindTurbineExtrLoad04, OFQ:RelRespFPSOEnvCont07, M-IH-Z:MultivarEnvContCvineCop16}. An extension of the direct sampling approach to three-dimensional problems was outlined in \cite{Vanem3Dcontour17}, and this method was applied to the tension in a mooring line of a semi-submersible in \cite{RMP:3DExtrValModTension19}. However, even though extensions of the direct sampling approach to environmental contour to higher dimensional problems is indeed possible, calculating the contours becomes increasingly cumbersome in higher dimensions.

\subsection{Contribution of this paper}

In this paper, an alternative way of constructing environmental contours is proposed, that easily generalises to arbitrary dimensions. With this method, environmental contours can be described as boundaries of Voronoi cells, which may easily be found from standard software packages at reasonable computational costs. The method makes use of Monte Carlo samples from the underlying distribution, but overcomes the common loop-problem of direct sampling methods, and can be used to produce convex contours with the desired probabilistic properties.

In Section 2 we briefly review the mathematical definition of environmental contours. 
In Section 3 we give a general introduction to Voronoi cells, before showing in Section 4 
that environmental contours may be interpreted as boundaries of Voronoi cells. 
In Section 5 we generalise results from Section 4 to the continuous limit, 
deriving additional theoretical insights, including an analytic formula for environmental contours 
in terms of a given percentile function.  
Section 6 details the practical application of the proposed algorithm, and examples in two and three dimensions 
are provided in Section 7. Some concluding remarks are provided in section 8. 
For brevity, proofs are contained in appendices.

\section{Definition of environmental contours} \label{sec:ECdef}
We consider a structure or component exposed to some environmental loads. 
The environmental loads can be represented by a vector of variables $\mathbf X \in \mathcal X \subseteq \mathbb{R}^n$, distributed according to some multivariate probability distribution $f_{\mathbf X}(\mathbf x)$.
We further define a  performance function $g(\mathbf x)$, where $\mathbf x$ is a specific environmental state, such that the structure or component remains intact/functioning as long as $g(\mathbf x)\geq 0$, and fails if $g(\mathbf x) < 0$. 

The failure region $\mathcal F = \{\mathbf x \in \mathcal X: g(\mathbf x) <0 \}$ and the  corresponding failure probability $p_f=P(\mathbf{X} \in \mathcal F)=\int_{\mathcal F} f_{\mathbf X}(\mathbf x) d \mathbf x$  are generally unknown. However, in many cases, one may argue based on physics that $\mathcal F$ must be convex.
Therefore, if we can find another convex set $\mathcal B$ such that $g(\mathbf x)\geq 0$ $\forall \mathbf x \in \mathcal B$, it follows from convexity theory that there exist a \emph{supporting hyperplane} $\Pi$ that separates  $\mathcal B$ and  $\mathcal F$  (i.e.  $\mathcal B \subseteq \Pi^{-}$ and $\mathcal F \subseteq \Pi^{+}$, where $\Pi^{-}$ and  $\Pi^{-}$ are the two half spaces separated by  $\Pi$), and  $p_f \leq P(\mathbf{X} \in \Pi^{+}) =\int_{\Pi^{+}} f_{\mathbf X}(\mathbf x) d \mathbf x $.

In particular, we may construct the set 
\begin{equation}
    \label{eq:ECdef}
    \mathcal B_{p_e} = \bigcap_{\mathbf u \in \mathcal U} \Pi_{p_e}^-(\mathbf u),
\end{equation}
where $\mathcal U$ denotes the set of all unit vectors in $\mathbb R^n$, i.e.
\begin{equation}
    \mathcal{U} = \{ \mathbf u \in \mathbb{R}^{n} \ | \ \norm{\mathbf u} = 1 \},
\end{equation}
and $\Pi_{p_e}^-(\mathbf u)$
is the half-space normal to $\mathbf u$ with the property that $P(\mathbf{X} \in \Pi^{-}(\mathbf u)) = 1-p_{e}$.
More precisely,
\begin{eqnarray}
    \label{eq:halfspace_1}
    \Pi_{p_e}^-(\mathbf u) = \{ \mathbf x : \mathbf u \cdot \mathbf x \leq C_{p_e}(\mathbf u) \},
\end{eqnarray}
where $C_{p_e}$ denotes the $p_e$-level percentile function, defined by
\begin{equation}
    \label{eq:C}
    C_{p_e}(\mathbf u) =  \textnormal{inf}\{ c: P(\mathbf u \cdot \mathbf X > c)\leq p_e\}.
\end{equation}
We will assume that the distribution of $\mathbf X$ is absolutely continuous 
with respect to the Lebesgue measure on $\mathbb{R}^{n}$, so 
the function $C_{p_e}(\mathbf u)$ in \eqref{eq:C} is well defined. 
We note also that \eqref{eq:ECdef} uniquely defines a convex set, as all half-spaces $\Pi_{p_e}^-(\mathbf u)$ are convex. 

Depending on the distribution of $\mathbf X$, the definition of $\mathcal B_{p_e}$ in \eqref{eq:ECdef} does not 
imply that all hyperplanes $\Pi_{p_e}(\mathbf u)$ intersect $\mathcal B_{p_e}$. (See for instance 
the discussion in Section \ref{sec:EC_as_VC} or the example given in Figure \ref{fig:micky_mouse}.) 
In the case where all hyperplanes $\Pi_{p_e}(\mathbf u)$ intersect $\mathcal B_{p_e}$,
the authors in \cite{Vanem:EnvCont14} state that $\mathbf X$ \emph{admits} a $p_{e}$-contour. We will make use of the 
equivalent definition below.

\begin{definition}
\label{def:proper_and_valid}
    Let $\mathcal B_{p_e}$ be a nonempty convex set in $\mathbb{R}^{n}$ and $p_e \in (0, 0.5)$. If 
    \begin{equation}
        P(\mathbf{X} \in \Pi^{+}) \leq p_{e}
        \label{eq:HS_prob}
    \end{equation}
    for any supporting half-space $\Pi^{+}$ of $\mathcal B_{p_e}$, we say that 
    $\partial \mathcal B_{p_e}$ is a \underline{valid} environmental contour of $\mathbf{X}$ with respect to the target 
    probability $p_{e}$. 
    If \eqref{eq:HS_prob} holds with equality for all the supporting half-spaces $\Pi^{+}$, 
    then $\partial \mathcal B_{p_e}$ is also a \underline{proper} environmental contour.
\end{definition}

In the case where a proper convex environmental contour exists, it is necessarily 
given by the representation in \eqref{eq:ECdef}. 
This follows from the fact that any closed convex subset  $\mathcal B \subset \mathbb{R}^{n}$ is the 
intersection of all supporting half-spaces that contain $\mathcal B$ 
(see e.g. Theorem 3.6.18 in \citep{Leonard:2015:Geometry_of_convex_sets}).
If all those half-spaces satisfy \eqref{eq:HS_prob} with equality, then the representation in \eqref{eq:ECdef} follows. 
For reference we state this in a separate proposition.

\begin{proposition}
	\label{prop:proper_cont}
    Assume that the random variable $\mathbf X$ admits a proper convex environmental contour
    $\partial \mathcal B_{p_e}$ with respect to a target probability $p_{e} \in (0, 0.5)$.
    Then the closure of $\mathcal B_{p_e}$ is uniquely defined by \eqref{eq:ECdef}.
\end{proposition}

In the following we will start by assuming that $\mathbf X$ admits a proper convex environmental contour, 
and also that the probabilities $P(\mathbf{X} \in \Pi^{+})$ can be computed without error. 
After introducing the connection with Voronoi cells and an algorithm for constructing $\mathcal B_{p_e}$, 
we present an approach that can be used when these assumptions are relaxed.  

\section{Voronoi cells}
The Voronoi diagram is a fundamental data structure in computational geometry that has found applications in 
a variety of fields, including physics, biology, cartography, crystallography, ecology, geology, anthropology, and 
meteorology to mention some \citep{Okabe:2000:Spatial_Tesselations}. Given a set of points ${p_{1}, \dots p_{k}}$ in a metric space $\mathcal{X}$, 
the Voronoi diagram is defined as the
partitioning of $\mathcal{X}$ into regions ${R_{1}, \dots R_{k}}$, such that $R_{i}$ contains all points in $\mathcal{X}$ whose 
distance to $p_{i}$ is not greater than their distance to any other $p_{j}$ for $j \neq i$. 
The region $R_{i}$ is often referred to as the \emph{Voronoi cell} of $p_{i}$ (with respect to the remaining points $p_{j}$, $j \neq i$).

\begin{figure}[h]
    \center{\includegraphics[width=0.9\textwidth]{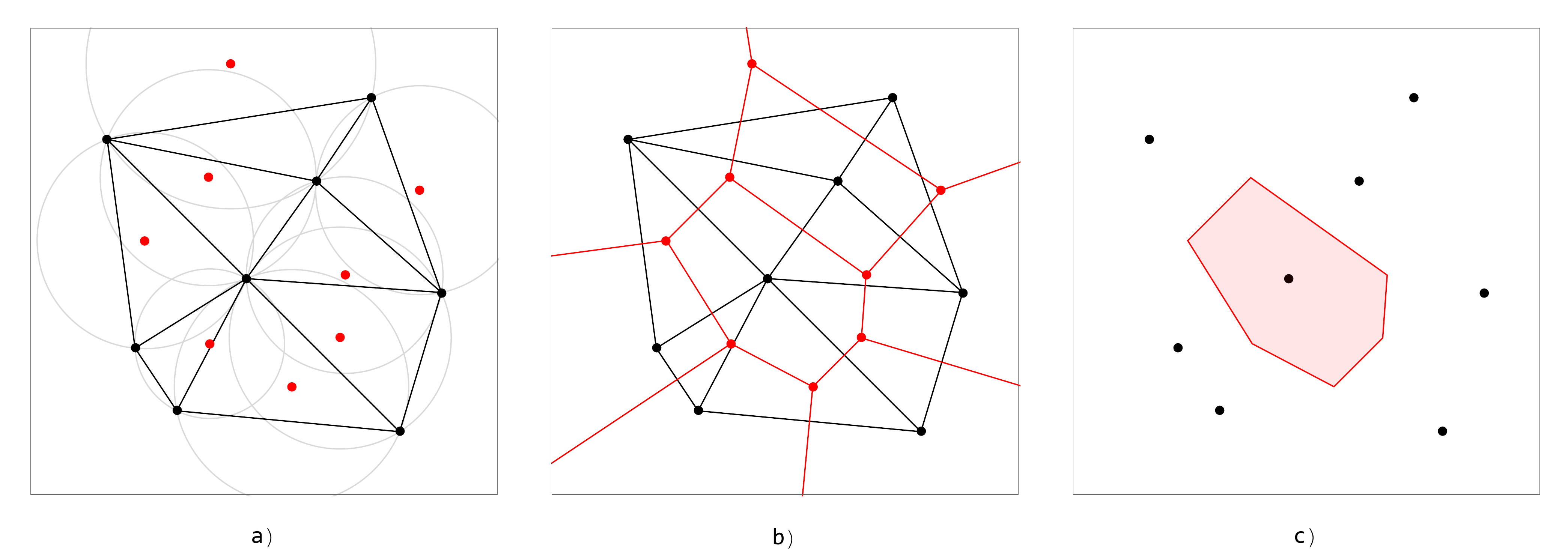}}
    \caption{Illustration of Delaunay triangulation and Voronoi diagram of a set of points. 
    a) A Delaunay triangulation of the 8 \textbf{black points} is defined as a triangulation such that no point 
    lies inside the circumcircle of any triangle. The {\color{red}{\textbf{red points}}} are the centers of each circumcircle.
    b) The Voronoi diagram ({\color{red}{\textbf{red lines}}}) corresponds to the graph with the circumcenters as edges. 
    c) The Voronoi cell of one of the points. }
    \label{fig:voronoi_delaunay}
\end{figure}

In its canonical form, a Voronoi diagram is constructed from a set of points in $\mathbb{R}^{n}$ endowed with the Euclidean metric,
and other alternatives are usually referred to as \emph{Generalised Voronoi diagrams} \cite{Schaller:2013:Set_Voronoi, Aurenhammer:1991:VDS}.
In this paper, we will consider the Voronoi cell of a point $\mathbf o \in \mathbb{R}^{n}$ with respect to a set $S \subset \mathbb{R}^{n}$.
We denote the Voronoi cell by $\vor{\mathbf o, S}$, and it is the set containing all points that are at least as close to $\mathbf o$ as any point in $S$,
measured by the Euclidean distance in $\mathbb{R}^{n}$.
\begin{equation}
    \label{eq:voronoi}
    \vor{\mathbf o, S} = \left\{ \mathbf x \in \mathbb{R}^{n} \ | \ \norm{\mathbf x - \mathbf o} \leq \inf_{\mathbf s \in S} \norm{\mathbf x - \mathbf s} \right\}.
\end{equation}
The distance function used to define $\vor{\mathbf o, S}$ could also be interpreted as the Hausdorff distance between the singleton set 
$\{ \mathbf o \}$ and $S$, but we will not make use of this property in this paper. To motivate the algorithm presented in this paper we will 
make use of the rather trivial property that if the set $S$ is finite, then it is equivalent to the canonical definition of (point) Voronoi cells 
as illustrated in Figure \ref{fig:voronoi_delaunay}. In the following section we show that an environmental contour can be represented 
as a Voronoi cell of the form \eqref{eq:voronoi}. A numerical approximation is then achieved by replacing the set $S$ in \eqref{eq:voronoi} 
with a finite subset, where available algorithms developed for canonical (point) Voronoi diagrams can be used. 
In this case we will also make use of the Delaunay triangulation of the finite point set, that correspond to the dual graph of the 
Voronoi diagram. This is illustrated for points in the plane in Figure \ref{fig:voronoi_delaunay}, and we refer to \citep{Okabe:2000:Spatial_Tesselations} 
for further details. 

\section{Environmental contours as boundaries of Voronoi cells}
\label{sec:EC_as_VC}
In this section we give a representation of the environmental contours described in Section \ref{sec:ECdef}
using Voronoi cells of the form \eqref{eq:voronoi}. 
We start by introducing the general construction and present some theoretical properties, 
in anticipation of a practical procedure for approximation of environmental contours that will follow in Section \ref{sec:algs}.

In Section \ref{sec:ECdef} we defined the environmental contours in terms of half-spaces that were parametrized by their perpendicular distance to the origin. 
However, a half-space may equivalently be parametrized in terms of perpendicular distance to any other point $\mathbf o \in \mathbb R^n$, i.e. 
\begin{eqnarray}
    \label{eq:halfspace_2}
    \Pi_{p_e}^-(\mathbf u) = \{ \mathbf x : \mathbf u \cdot (\mathbf x - \mathbf o ) \leq C^{\mathbf o}_{p_e}(\mathbf u) \},
\end{eqnarray}
with 
\begin{eqnarray}
    \label{eq:C_p_o}
    C^{\mathbf o}_{p_e}(\mathbf u) =  \textnormal{inf}\{ c: P(\mathbf u \cdot (\mathbf X -\mathbf o)> c)\leq p_e\}.
\end{eqnarray}
By comparing \eqref{eq:ECdef} and \eqref{eq:C_p_o} it is evident that
\begin{eqnarray}
    \label{eq:C_conversion}
    C^{\mathbf o}_{p_e}(\mathbf u) = C_{p_e}(\mathbf u) - \mathbf u \cdot \mathbf o,
\end{eqnarray}
and that the two definitions of $\Pi_{p_e}^-(\mathbf u)$ given in \eqref{eq:halfspace_1} and \eqref{eq:halfspace_2} are equivalent.

Using this alternative parametrization for $\Pi_{p_e}^-(\mathbf u)$, we define the set $\mathcal S^{\mathbf o}_{p_e}(U)$ as
\begin{equation}
    \label{eq:S}
    \mathcal S^{\mathbf o}_{p_e}(U) =  \{ \mathbf s_{p_e}^{\mathbf o,\mathbf u} =\mathbf o + 2C^{\mathbf o}_{p_e}(\mathbf u) \mathbf u\}_{ \mathbf u \in U},
\end{equation}
where $U$ is a subset of the unit vectors in $\mathbb{R}^{n}$.

A point $\mathbf s_{p_e}^{\mathbf o,\mathbf u} \in \mathcal S^{\mathbf o}_{p_e}(U)$ represent the reflection of the point $\mathbf o\in\mathbb R^n$ with respect to the boundary of the half-space $\Pi_{p_e}^-(\mathbf u)$ (i.e. with respect to $\Pi_{p_e}(\mathbf u)$). 
Stated differently, the half-space  
$\Pi_{p_e}^-(\mathbf u)$ contains all points that are closer to $\mathbf o$ than to $\mathbf s_{p_e}^{\mathbf o,\mathbf u}$. 
Intuitively, if $\mathbf o$ is in the interior of $\mathcal B_{p_e}$,
then all points in the convex set $\mathcal B_{p_e}$ should be closer to 
$\mathbf o$ than to any point in $\mathcal S^{\mathbf o}_{p_e}(U)$.
This means that $\mathcal B_{p_e}$ is the Voronoi cell of $\mathbf o$ with respect to the set of points $\mathcal S^{\mathbf o}_{p_e}(U)$.
The latter insight is stated formally as a lemma below.

\begin{lemma}
    \label{lemma:cont_vor}
    Let $\mathcal{B}_{p_e}$ be defined as in \eqref{eq:ECdef}. Then
    \begin{align*}
        \mathbf o \in \mathcal B_{p_e} &\Longleftrightarrow C^{\mathbf o}_{p_e}(\mathbf u) \geq 0 \ \forall \ \mathbf u \in \mathcal{U}, \\
        \mathbf o \in \mathcal B_{p_e} &\setminus \partial \mathcal B_{p_e} \Longleftrightarrow C^{\mathbf o}_{p_e}(\mathbf u) > 0 \ \forall \ \mathbf u \in \mathcal{U}.    
    \end{align*}
    Furthermore, if $\mathbf o \in \mathcal B_{p_e} \setminus \partial \mathcal B_{p_e}$ we have for any subset $U \subseteq \mathcal{U}$ that 
    \begin{equation*}
        \vor{\mathbf o, \mathcal S^{\mathbf o}_{p_e}(U)} = \bigcap_{\mathbf u \in U} \Pi_{p_e}^-(\mathbf u),
    \end{equation*}
    where $\vor{\cdot, \cdot}$ is the Voronoi cell as defined in \eqref{eq:voronoi}.
\end{lemma}

The proof is given in Appendix \ref{app:proof_lemma_cont_vor}. 
Using this result we arrive at the following proposition that 
motivates the algorithm presented in this paper. 

\begin{proposition}
    \label{prop:main}
    Let $\mathcal{B}_{p_e}$ be defined as in \eqref{eq:ECdef},
    and let $U_{1}$ and $U_{2}$ be sets of unit vectors in 
    $\mathbb{R}^{n}$, 
    such that $U_{1} \subseteq U_{2} \subseteq \mathcal{U}$. 
    If $\mathbf o \in \mathcal B_{p_e} \setminus \partial \mathcal B_{p_e}$ then the following holds:

    \begin{equation*}
        \mathcal B_{p_e} = \vor{\mathbf o, \mathcal S^{\mathbf o}_{p_e}(\mathcal{U})} 
        \subseteq \vor{\mathbf o, \mathcal S^{\mathbf o}_{p_e}(U_{2})}
        \subseteq \vor{\mathbf o, \mathcal S^{\mathbf o}_{p_e}(U_{1})}.
    \end{equation*}

\end{proposition}
\vspace{0.05cm}

\begin{figure}[h!]
    \center{\includegraphics[width=0.5\textwidth]{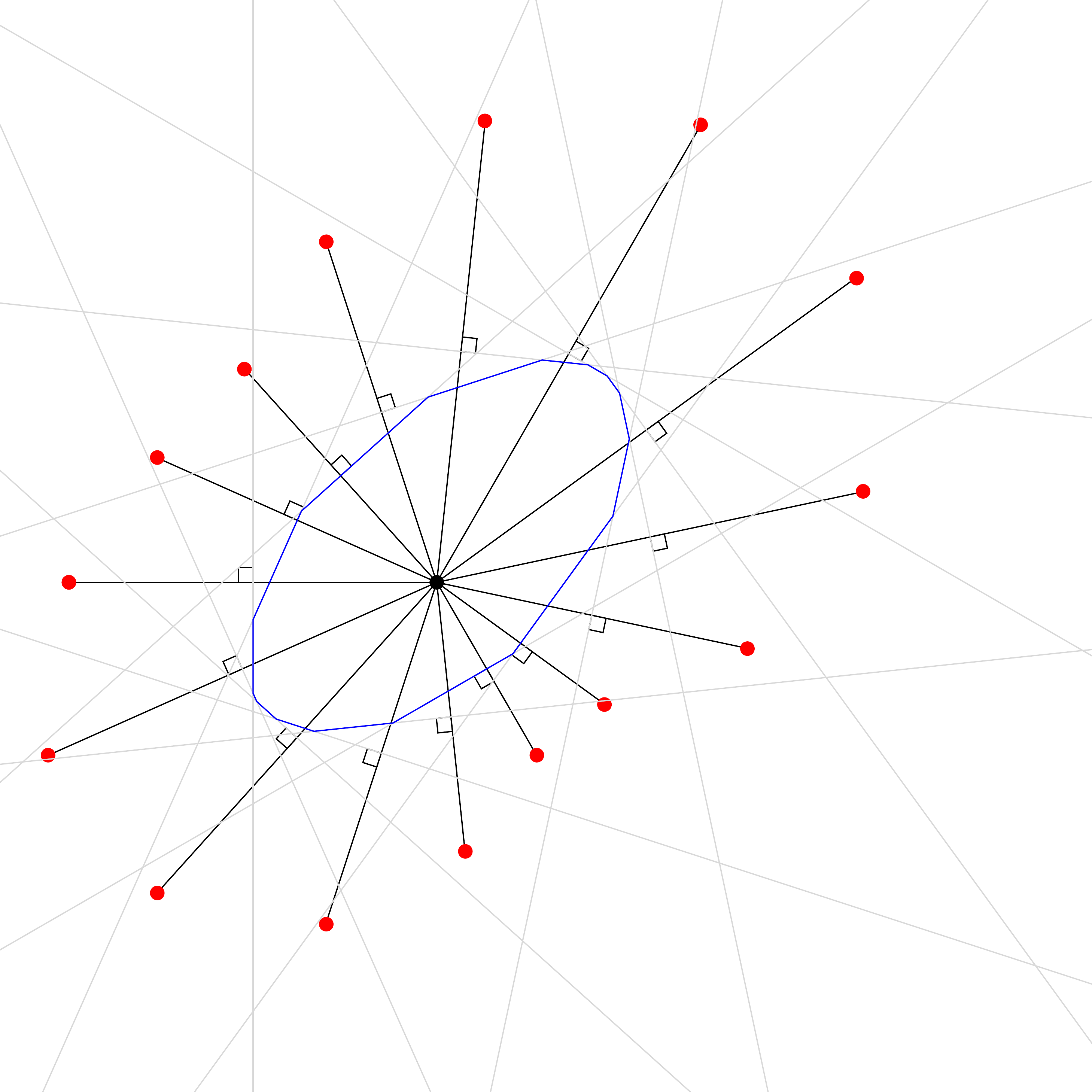}}
    \caption{Construction of environmental contour using the Voronoi method. 
    The black point is the chosen origin $\mathbf o \in \mathcal B_{p_e} \setminus \partial \mathcal B_{p_e}$. 
    The red points correspond to the finite set $\mathcal S^{\mathbf o}_{p_e}$. 
    The boundaries of the half planes $\Pi_{p_e}^-(\mathbf u)$ half way between $\mathbf o$ and the respective 
    points $\mathbf s_{p_e}^{\mathbf o,\mathbf u}\in \mathcal S^{\mathbf o}_{p_e}$ are drawn as light grey lines, 
    and their perpendicularity on the black lines from  $\mathbf o$ to $\mathbf s_{p_e}^{\mathbf o,\mathbf u}$ is 
    indicated with small squares. The boundary of the Voronoi cell of $\mathbf o$ with respect to 
    $\mathcal S^{\mathbf o}_{p_e}$ is drawn in blue, and it can be seen that the grey lines are tangential on the Voronoi cell.}
    \label{fig:vor_alg_illustration}
\end{figure}

\noindent This proposition follows directly from Lemma \ref{lemma:cont_vor}
(see Appendix \ref{app:proof_prop_main} for details).
The first interesting observation is that the environmental contour, $\partial \mathcal B_{p_e}$, 
can be represented as the boundary of the Voronoi cell $\vor{\mathbf o, \mathcal S^{\mathbf o}_{p_e}(\mathcal{U})}$.
This insight immediately suggests a new algorithm for numerical approximation of environmental contours,
by replacing the set of unit vectors $\mathcal{U}$ with a finite subset
$U = \{ \mathbf u_{i} \ | \ \mathbf u_{i} \in \mathcal{U}, i = 1, \dots, k  \}$, as illustrated in Figure \ref{fig:vor_alg_illustration}. 
The proposition also states that any such approximation of a \emph{proper} convex environmental contour will be conservative, 
in the sense that the resulting Voronoi cell is guaranteed to contain $\mathcal B_{p_e}$. 
Accordingly, any approximation will be a \emph{valid} convex environmental contour. 
Moreover, including more unit vectors in the set $U$ improves the approximation (or at least does not make it worse).
Intuitively, the error in the approximation can be made arbitrarily small, although this 
naturally will depend on the sampling strategy used. 

A natural procedure for approximating $\mathcal B_{p_e}$ could therefore be as follows:

\begin{center}
    \begin{minipage}{0.8\textwidth}
        \begin{itemize}
            \item[\textbf{Step 1}] Select a set of unit vectors $U = \{ \mathbf{u}_{j}\}^M_{j=1}$.
            \item[\textbf{Step 2}] Compute $C_{p_e}(\mathbf{u}_{1}), \dots, C_{p_e}(\mathbf{u}_{M})$.
            \item[\textbf{Step 3}] Compute $\mathcal S^{\mathbf o}_{p_e}(U)$ for some $\mathbf o \in \mathcal B_{p_e} \setminus \partial \mathcal B_{p_e}$.
            \item[\textbf{Step 4}] Compute the Voronoi cell of $\mathbf o$ with respect to $\mathcal S^{\mathbf o}_{p_e}(U)$. 
        \end{itemize}
    \end{minipage}
\end{center}

Under the assumption that a proper convex environmental contour exists (for the given random variable
$\mathbf X$ and target probability $p_{e}$), the set
$\widehat{\mathcal B}_{p_e} = \vor{\mathbf o, \mathcal S^{\mathbf o}_{p_e}(U)}$ is guaranteed 
to contain $\mathcal B_{p_e}$, and the difference can be made arbitrarily small by 
including sufficiently many unit vectors in $U$. 
For practical application, however, it is not reasonable to assume that the function $C_{p_e}(\mathbf{u})$ 
can be computed exactly, and we might not have \emph{a priori} a point $\mathbf o \in \mathcal B_{p_e} \setminus \partial \mathcal B_{p_e}$.
We will postpone these questions to Section \ref{sec:algs}.
For now, we will assume that a point $\mathbf o \in \mathcal B_{p_e} \setminus \partial \mathcal B_{p_e}$ is given and 
that the function $C_{p_e}(\mathbf{u})$ can be evaluated without error, in order to study the 
final major assumption. Namely, that the random variable of interest $\mathbf X$
admits a proper convex environmental contour for the target probability $p_{e}$.

\begin{figure}[h!]
    \center{\includegraphics[width=0.9\textwidth]{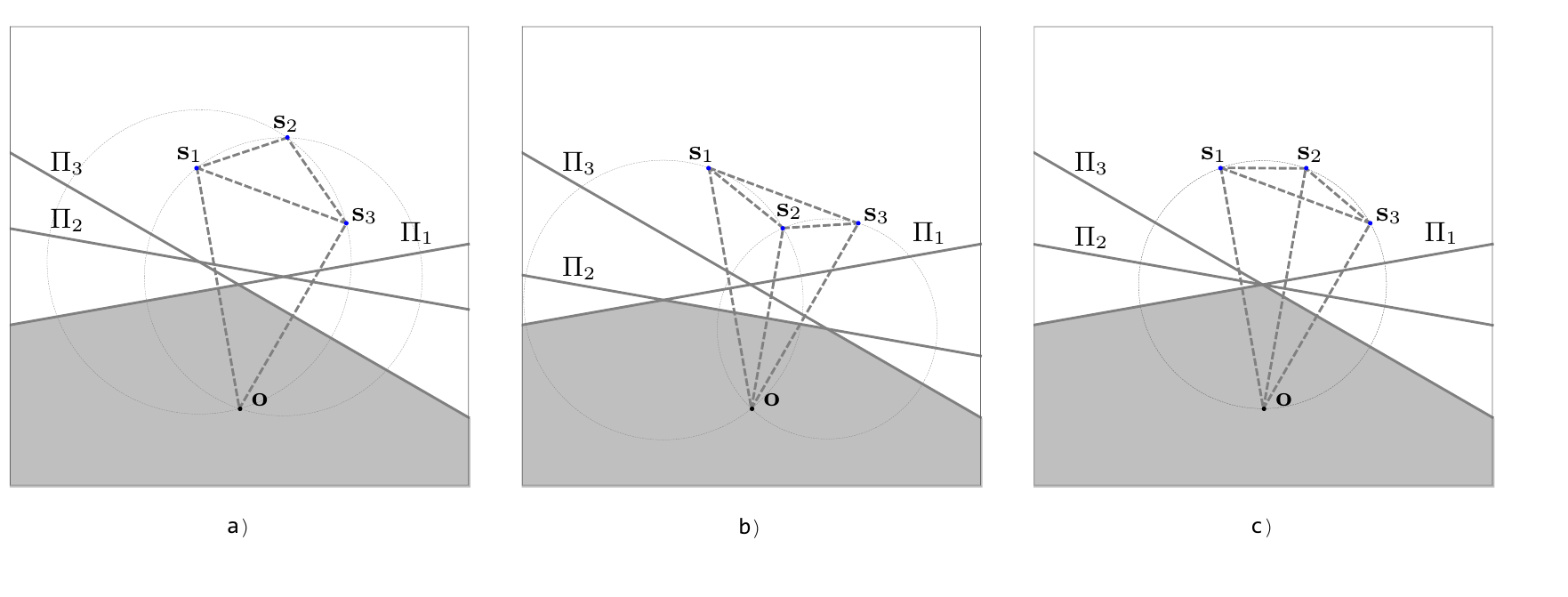}}
    \caption{Three points from $\mathcal S^{\mathbf o}_{p_e}(U)$ with corresponding hyperplanes, 
    $\mathbf{s}_{i} = \mathbf s_{p_e}^{\mathbf o,\mathbf{u}_{i}}$ and $\Pi_{i} = \Pi_{p_e}(\mathbf{u}_{i})$ for three unit vectors 
    $U = \{\mathbf{u}_{1}, \mathbf{u}_{2}, \mathbf{u}_{3} \}$. The Voronoi cell $\vor{\mathbf o, \mathcal S^{\mathbf o}_{p_e}(U)}$
    corresponds to the shaded area in each figure, and the dual Delaunay triangulation is indicated with dashed lines.  
    a) $\Pi_{2}$ is not a supporting hyperplane of $\vor{\mathbf o, \mathcal S^{\mathbf o}_{p_e}(U)}$ since 
    $\mathbf{s}_{2}$ is not connected to $\mathbf{o}$ by any Delaunay edge.
    b) All planes $\Pi_{i}$ intersect $\vor{\mathbf o, \mathcal S^{\mathbf o}_{p_e}(U)}$ as $\mathbf{s}_{i}$ is connected to $\mathbf{o}$ by a Delaunay edge for all $i$.
    c) The Delaunay triangulation is not unique, and $\Pi_{2}$ only intersects a vertex of $\vor{\mathbf o, \mathcal S^{\mathbf o}_{p_e}(U)}$.
    }
    \label{fig:three_planes_vor}
\end{figure}

In practice, it might not be possible to determine \emph{a priori} whether 
a proper convex environmental contour exists. 
To see how we might account for this issue, we first study what will happen if 
$\mathbf X$ \emph{does not} admit a proper convex environmental contour. 
In Figure \ref{fig:three_planes_vor} we reproduce the example given in \cite{Vanem:EnvCont14}, illustrating 
the scenario where a supporting half-space can have exceedance probability larger than $p_{e}$. 
That is, one of the hyperplanes $\Pi_{p_e}^{-}(\mathbf u)$ in \eqref{eq:ECdef} does not intersect $\mathcal B_{p_e}$.
Hence, if a scenario such as the one in Figure \ref{fig:three_planes_vor} a) occur, this means that a 
proper environmental contour cannot exist (for the selected target probability $p_{e}$). 
As we illustrate in the figure, there is an interesting connection with the dual representation of the Voronoi cell, the 
Delaunay triangulation, that can be exploited when studying this problem. 
We recall that every edge on a Voronoi cell corresponds to the circumcenter of a Delaunay triangle (in general a Delaunay simplex for higher dimensions), 
and we say that a Delaunay triangulation connects two points $\mathbf{a}, \mathbf{b} \in \mathcal{X}$ if both 
$\mathbf{a}$ and $\mathbf{b}$ are part of the same triangle (simplex) in the triangulation. With this terminology, we may state 
the observation made in Figure \ref{fig:three_planes_vor} formally as follows.

\begin{proposition}
    \label{prop:delaunay_proper_EC}
    Assume $\partial \mathcal{B}_{p_e}$ is a proper convex environmental contour with $\mathcal{B}_{p_e}$ defined as in \eqref{eq:ECdef}. 
    Let $\mathcal S^{\mathbf o}_{p_e}(U)$ be defined as in \eqref{eq:S} for some finite set 
    $U \subset \mathcal{U}$,
    and $\mathbf o \in \mathcal B_{p_e} \setminus \partial \mathcal B_{p_e}$.

    Then, for all $\mathbf s \in \mathcal S^{\mathbf o}_{p_e}(U)$, there exists a Delaunay triangulation of the point set 
    $\{ \mathbf o \} \cup S^{\mathbf o}_{p_e}(U)$ that connects $\mathbf s$ and $\mathbf o$.
\end{proposition}

A proof of Proposition \ref{prop:delaunay_proper_EC} is given in Appendix \ref{app:proof_prop_delaunay}, 
where we refer to \citep{Okabe:2000:Spatial_Tesselations} for results regarding the Voronoi-Delaunay duality. 
We may also make use of the fact that a Delaunay triangulation of a point set is unique if the points are in \emph{general position}. 
In the general $n$-dimensional case, a set $\mathbf P$ of points is in \emph{general position} if the affine hull of 
$\mathbf P$ is $n$-dimensional, and there is no subset of $n+2$ points in $\mathbf P$ that lie on the boundary of a ball 
whose interior does not intersect $\mathbf P$.
Figure \ref{fig:three_planes_vor} c) shows a scenario where this condition is violated. Here, 
the affine hull of the set $\mathbf P = \{ \mathbf o, \mathbf{s}_{1}, \mathbf{s}_{2}, \mathbf{s}_{3} \}$ is clearly $2$-dimensional, 
but the four points in $\mathbf P$ all lie on a circle (whose interior does not contain any points in $\mathbf P$). 
Hence, the Delaunay triangulation is not unique. There are in fact two possible Delaunay triangulations as illustrated in Figure \ref{fig:three_planes_vor} c),
$\{ \{ \mathbf o, \mathbf{s}_{1}, \mathbf{s}_{3} \}, \{ \mathbf{s}_{1}, \mathbf{s}_{2}, \mathbf{s}_{3} \} \}$
and $\{ \{ \mathbf o, \mathbf{s}_{1}, \mathbf{s}_{2} \}, \{ \mathbf{o}, \mathbf{s}_{2}, \mathbf{s}_{3} \} \}$.
Using this condition for uniqueness together with Proposition \ref{prop:delaunay_proper_EC}, we immediately achieve the following 
convenient result.

\begin{corollary}
    \label{cor:delaunay_proper_EC}
    Under the assumptions of Proposition \ref{prop:delaunay_proper_EC}, if also the points in $\{ \mathbf o \} \cup S^{\mathbf o}_{p_e}(U)$
    are in general position, then the Delaunay triangulation is unique and connects all points 
    $\mathbf s \in \mathcal S^{\mathbf o}_{p_e}(U)$ with $\mathbf o$.
\end{corollary}

\begin{figure}[h!]
    \center{\includegraphics[width=0.4\textwidth]{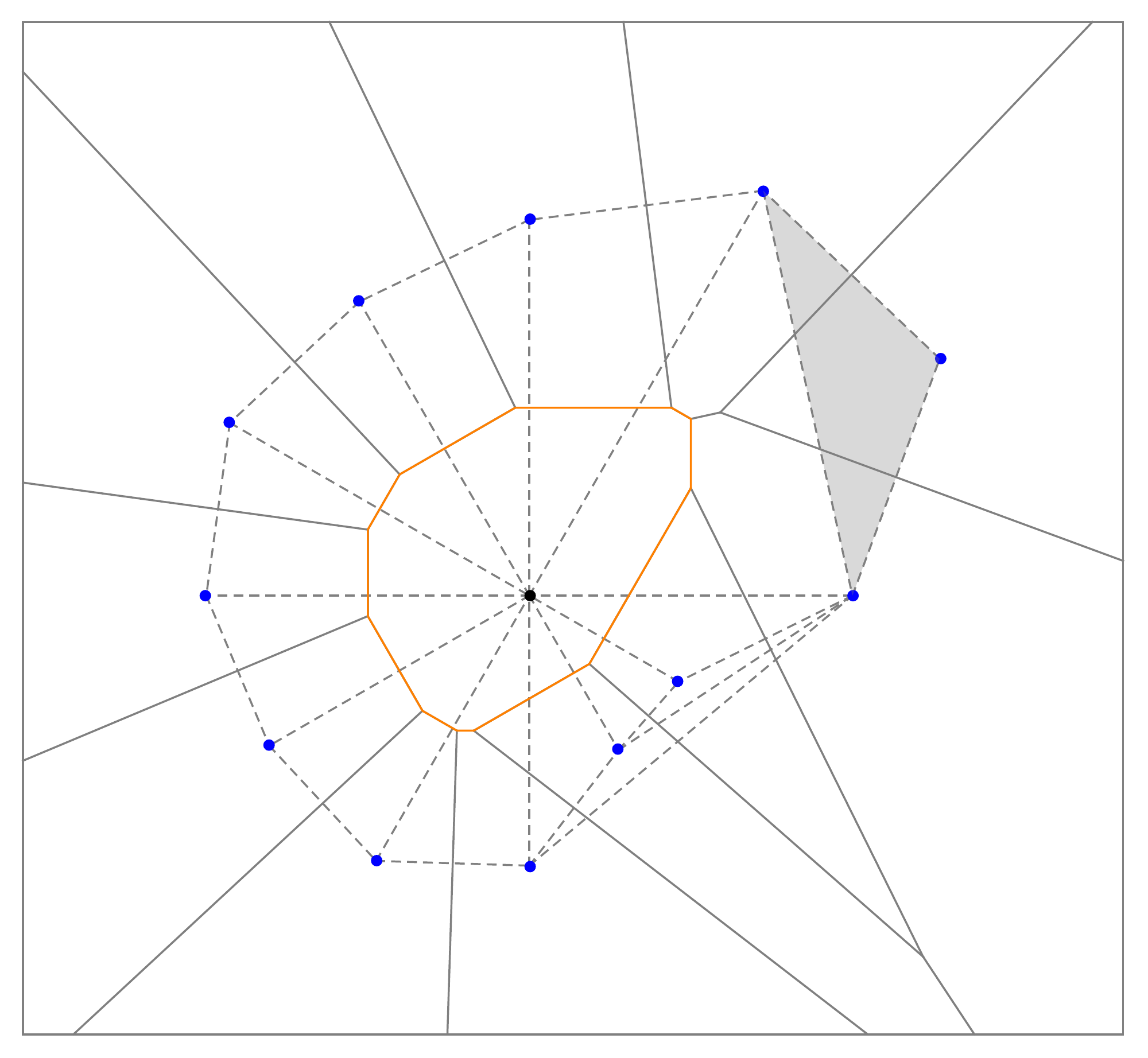}}
    \caption{Illustration of the idea behind Proposition \ref{prop:delaunay_proper_EC} and Corollary \ref{cor:delaunay_proper_EC} in 2D.
    The dashed lines shows the Delaunay triangulation of the points $\{ \mathbf o \} \cup S^{\mathbf o}_{p_e}(U)$, which are in general position. 
    The grey triangle contains a point $\mathbf s \in S^{\mathbf o}_{p_e}(U)$ that is not connected to $\mathbf o$. Hence, 
    no proper convex environmental contour exists for the selected probability $p_{e}$ and the random variable $\mathbf X$ used 
    to generate $S^{\mathbf o}_{p_e}(U)$.
    }
    \label{fig:illustration_del_property}
\end{figure}

Corollary \ref{cor:delaunay_proper_EC} is useful as it gives a criterion for checking whether a proper convex environmental contour exists, 
and for identification of directions (for which unit vector $\mathbf u$) there might be problems. 
The general idea is also illustrated in Figure \ref{fig:illustration_del_property}, where we can conclude that no proper convex environmental contour 
exists, for the given distribution of $\mathbf X$ and target probability $p_{e}$, as the grey shaded triangle contains a point $\mathbf s \in S^{\mathbf o}_{p_e}(U)$ which is not connected with $\mathbf o$.

\section{Voronoi contours in the continuous limit}
\label{sec:continious_limit}

\begin{figure}[h!]
    \center{\includegraphics[width=0.45\textwidth]{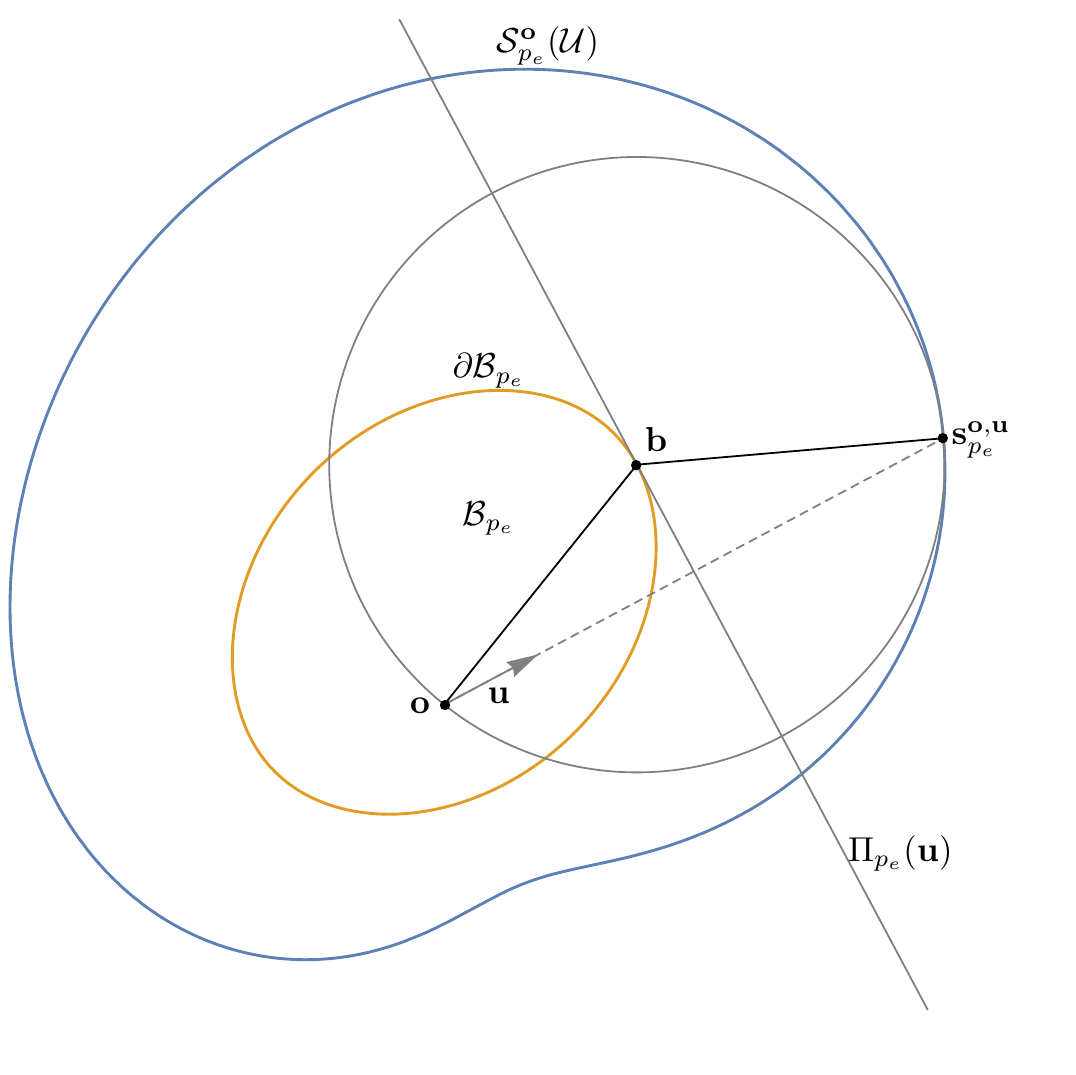}}
    \caption{Geometric illustration of Proposition \ref{prop:sphere_property} in 2D. 
    For any $\mathbf{u} \in \mathcal{U}$ there exists some $\mathbf b \in \Pi_{p_e}(\mathbf{u}) \cap \partial \mathcal{B}_{p_e}$, such that the circle centered at $\mathbf b$ that also contains $\mathbf o$ is tangent to 
    $\mathcal S^{\mathbf o}_{p_e}(\mathcal{U})$ at $\mathbf s_{p_e}^{\mathbf o,\mathbf{u}}$, and contains no points from $\mathcal S^{\mathbf o}_{p_e}(\mathcal{U})$ in its interior.
    }
    \label{fig:ball_property}
\end{figure}

From the illustrations in Figure \ref{fig:three_planes_vor} and Figure \ref{fig:illustration_del_property}, we could 
also imagine what happens as more points are added, moving to the limit as $S^{\mathbf o}_{p_e}(U) \rightarrow S^{\mathbf o}_{p_e}(\mathcal{U})$.
Consider the Delaunay triangle $\{ \mathbf o, \mathbf{s}_{2}, \mathbf{s}_{3} \}$ in Figure \ref{fig:three_planes_vor} b). 
This triangle has the property that its circumcircle contains no other points from $S^{\mathbf o}_{p_e}(U)$ in its interior. 
As the points $\mathbf{s}_{2}$ and $\mathbf{s}_{3}$ move arbitrarily close together, the circumcircle of this "triangle" is the 
circle that contain $\mathbf o$ and is tangential to $\mathbf{s}_{2} \approx \mathbf{s}_{3}$. Moreover, the center of this circle is a point on
$\partial \mathcal{B}_{p_e}$. From this intuition we arrive at the geometric property of proper convex environmental contours,
which is illustrated in Figure \ref{fig:ball_property}. We state this formally in Proposition \ref{prop:sphere_property}, with a proof given in 
Appendix \ref{app:proof_prop_sphere_property}. 

\begin{proposition}
    \label{prop:sphere_property}
    Assume $\partial \mathcal{B}_{p_e}$ is a proper convex environmental contour with $\mathcal{B}_{p_e}$ defined as in \eqref{eq:ECdef}.
    Let $\mathcal S^{\mathbf o}_{p_e}(\mathcal{U})$ be as in \eqref{eq:S} 
    and define, for any $\mathbf b \in \partial \mathcal{B}_{p_e}$ and $\mathbf o \in \mathcal B_{p_e} \setminus \partial \mathcal B_{p_e}$, 
    the $n$-dimensional ball $\mathcal{W}^{\mathbf o}(\mathbf b) = \{ \mathbf x \in \mathbb{R}^{n} \ | \ \norm{\mathbf x - \mathbf b} \leq \norm{\mathbf b - \mathbf o} \}$. 
    
    Then for any $\mathbf{u} \in \mathcal{U}$, there exists some $\mathbf b \in \Pi_{p_e}(\mathbf{u}) \cap \partial \mathcal{B}_{p_e}$ where 
    $\mathcal S^{\mathbf o}_{p_e}(\mathcal{U}) \cap \mathcal{W}^{\mathbf o}(\mathbf b) \subseteq \partial \mathcal{W}^{\mathbf o}(\mathbf b)$, 
    and $\mathbf s_{p_e}^{\mathbf o,\mathbf{u}} \in \partial \mathcal{W}^{\mathbf o}(\mathbf b)$.
\end{proposition}

A consequence of the geometric property stated in Proposition \ref{prop:sphere_property}
is that, given a parametrization of unit vectors in $\mathbb{R}^{n}$, 
we will be able to derive a parametric characterization of $\partial \mathcal{B}_{p_e}$.
The key insight from Figure \ref{fig:ball_property} is that, given certain regularity assumptions,
the vectors tangential to the set $\mathcal S^{\mathbf o}_{p_e}(\mathcal{U})$ and the 
ball $\mathcal{W}^{\mathbf o}(\mathbf b)$ coincide at $\mathbf s_{p_e}^{\mathbf o,\mathbf{u}}$. 
This will eventually let us derive a parametric representation of 
the set $\partial \mathcal{B}_{p_e}$ as a $(n-1)$-dimensional manifold. 
So now, motivated by the properties derived in the discrete scenario using tools from computational geometry,
i.e. the Voronoi and Delaunay tessellations, we will move to the continuous limit 
and study environmental contours in the context of differential geometry. 

We will start by assuming that the set $\mathcal S^{\mathbf o}_{p_e}(\mathcal{U})$, 
viewed as a $(n-1)$-dimensional manifold embedded in $\mathbb{R}^{n}$, is differentiable. 
We recall that a $m$-dimensional manifold $\mathcal S$ in $\mathbb{R}^{n}$, for $m \leq n$, 
can be represented by a set of \emph{charts} $\sigma_{i} : V_{i} \rightarrow \mathcal S$, 
where $V_{i}$ are open non-empty subsets of $\mathbb{R}^{m}$.
Any set of charts 
$\{\sigma_{i}, V_{i} \}_{i}$ that cover $\mathcal S$, i.e. $\mathcal S = \cup_{i} \sigma_{i}(V_{i})$, is called an \emph{atlas} of $\mathcal S$.
We will in particular consider a regular parametrization of the unit $(n-1)$-sphere $\mathcal U$, 
by which we mean a set of charts $\{\sigma_{i}, V_{i} \}_{i}$ covering $\mathcal U$
where each $\sigma_{i}$ is smooth 
and where the Jacobi matrix of $\sigma_{i}$ has rank $n-1$ at any point in $V_{i}$.
With the canonical alternative of spherical coordinates in mind, 
we will let $\{\mathbf{u}_{i}( \bm{\theta}) \ | \ \bm{\theta} \in \Theta_{i} \}_{i}$
denote an atlas of $\mathcal U$ with these properties.
With some abuse of terminology, we will also refer to $\{\mathbf{u}_{i}( \bm{\theta}) \ | \ \bm{\theta} \in \Theta_{i} \}_{i}$ 
as a \emph{regular parametrization} of $\mathcal U$.
Given such a regular parametrization of $\mathcal U$, we will continue to construct 
corresponding parametrizations of $\mathcal S^{\mathbf o}_{p_e}(\mathcal{U})$
and eventually $\partial \mathcal{B}_{p_e}$. But first we will need a preliminary result given in Lemma \ref{lemma:au} below.

\begin{lemma}
    \label{lemma:au}
    Assume $\partial \mathcal{B}_{p_e}$ is a proper convex environmental contour with $\mathcal{B}_{p_e}$ defined as in \eqref{eq:ECdef}, 
    let $\mathbf o \in \mathcal B_{p_e} \setminus \partial \mathcal B_{p_e}$ and 
    assume $\mathcal S^{\mathbf o}_{p_e}(\mathcal{U})$ is a differentiable manifold.\\

    If the pair $(\mathbf{a}, \mathbf{u})$, for some $\mathbf{a} \in \mathbb{R}^{n}$ and $\mathbf{u} \in \mathcal{U}$, 
    satisfies the following
    \begin{center}
        \begin{minipage}{0.55\textwidth}
            \begin{enumerate}    
                \item $\norm{\mathbf{a} - \mathbf{o}} = \norm{\mathbf s_{p_e}^{\mathbf o,\mathbf u} - \mathbf{a}}$, and
                \item $(\mathbf s_{p_e}^{\mathbf o,\mathbf u} - \mathbf{a})$ is orthogonal to $\mathcal S^{\mathbf o}_{p_e}(\mathcal{U})$ at $\mathbf s_{p_e}^{\mathbf o,\mathbf u}$,
            \end{enumerate}
        \end{minipage}
    \end{center}
    then $\{ \mathbf{a} \} = \Pi_{p_e}(\mathbf{u}) \cap \partial \mathcal B_{p_e}$.
\end{lemma}

In the proof of Lemma \ref{lemma:au}, given in Appendix \ref{app:proof_lemma_au}, we also show that 
for any $\mathbf u \in \mathcal{U}$, $\Pi_{p_e}(\mathbf{u}) \cap \partial \mathcal B_{p_e}$ is a singleton set, 
as $\P_{p_e}i(\mathbf{u}) \cap \partial \mathcal B_{p_e}$ is nonempty when $\partial \mathcal{B}_{p_e}$ is a proper convex environmental contour
and the pair $(\mathbf b, \mathbf u)$ satisfies the conditions in Lemma \ref{lemma:au} for any $\mathbf{b} \in \Pi_{p_e}(\mathbf{u}) \cap \partial \mathcal B_{p_e}$.
This means that the set $\mathcal{B}_{p_e}$ has no "flat parts", and that $\mathcal{B}_{p_e}$ is in fact strictly convex. 
But besides this, the conditions in Lemma \ref{lemma:au} will also serve as a more practical 
criterion to verify that a given mapping (soon to be given explicitly) 
gives a representation of the environmental contour $\partial \mathcal{B}_{p_e}$. 
This result is summarised in Proposition \ref{prop:map_to_dB_criterion} below, with a proof given in Appendix \ref{app:proof_prop_map_to_dB_criterion}.

\begin{proposition}
    \label{prop:map_to_dB_criterion}
    Let $F : \mathcal{U} \rightarrow \mathbb{R}^{n}$ be a mapping such that the assumptions and conditions 
    of Lemma \ref{lemma:au} hold for any pair $(F(\mathbf u), \mathbf u)$. Then $F(\mathcal{U}) = \partial \mathcal{B}_{p_e}$.
\end{proposition}

Now, the next step is to introduce a specific parametrization of $\partial \mathcal{B}_{p_e}$
that we will use Proposition \ref{prop:map_to_dB_criterion} to verify. 
We will achieve this by mapping a parametrization of the unit $(n-1)$-sphere $\mathcal U$ to 
a parametrization of $\partial \mathcal{B}_{p_e}$.
This idea has been explored in \cite{Vanem:EnvCont14, Huseby:stk4400} 
for the $2$-dimensional case using the parametrization $\mathbf{u}(\theta) = (\cos (\theta), \sin (\theta))$, 
where also the existence of a proper convex environmental contour is determined from properties related to the 
parametrized percentile function $C_{p_{e}}(\theta) = C_{p_{e}}(\mathbf{u}(\theta))$.
In the following we will extend this to the $n$-dimensional case.

Let $\{\mathbf{u}_{i}( \bm{\theta}) \ | \ \bm{\theta} \in \Theta_{i} \}_{i}$ be the regular 
parametrization of $\mathcal U$ introduced previously. 
Suppressing the index $i$, for any chart $\mathbf{u} (\bm{\theta}) : \Theta \rightarrow \mathcal U$
we define
the functions $C^{\mathbf o}_{p_e}(\bm{\theta})$ and $\mathbf s_{p_e}^{\mathbf o}(\bm{\theta})$ accordingly,
\begin{equation}
    \begin{split}
        C^{\mathbf o}_{p_e}(\bm{\theta}) &= C^{\mathbf o}_{p_e}(\mathbf u(\bm{\theta})) : \Theta \rightarrow \mathbb{R}, \\
        \mathbf s_{p_e}^{\mathbf o}(\bm{\theta}) &= \mathbf o + 2C^{\mathbf o}_{p_e}(\bm{\theta}) \mathbf u(\bm{\theta}) : \Theta \rightarrow \mathbb{R}^{n}, 
    \end{split}
\end{equation}
where we will assume that both $\mathbf{u}( \bm{\theta})$ and $C^{\mathbf o}_{p_e}(\bm{\theta})$ are 
continuously differentiable as functions of $\bm{\theta}$, 
and let $\nabla_{\bm{\theta}}$ denote the Jacobian. 
That is, for functions $\mathbf{f}: \Theta \rightarrow \mathbb{R}^{m}$,  
$\nabla_{\bm{\theta}} \mathbf{f}$ is the $m \times (n-1)$ matrix with entries 
$[\nabla_{\bm{\theta}} \mathbf{f}]_{i, j} = \partial \mathbf{f}_{i} / \partial \bm{\theta}_{j}$.
The assumption that $\mathbf{u}(\bm{\theta})$ is a \emph{regular} parametrization means that 
we also assume that $\nabla_{\bm{\theta}} \mathbf{u}(\bm{\theta})$ has rank $n-1$ for any $\bm{\theta} \in \Theta$.

\begin{theorem}[Representation of proper convex environmental contours]
    \label{thm:param}
    Assume the $n$-dimensional random variable $\mathbf X$ admits a proper convex environmental contour
    $\partial \mathcal B_{p_e}$ with respect to a target probability $p_{e} \in (0, 0.5)$, 
    and assume that the $p_e$-level percentile function $C_{p_{e}}(\textbf u)$ is $k$-times continuously differentiable 
    on the unit $(n-1)$-sphere for $k \geq 1$.

    Then $\mathcal{B}_{p_e}$ is strictly convex, and $\partial \mathcal B_{p_e}$ is a $(k-1)$-times differentiable manifold.
    Furthermore, if  $\{\mathbf{u}_{i}( \bm{\theta}) \ | \ \bm{\theta} \in \Theta_{i} \}_{i = 1}^m$ is a regular parametrization of the unit $(n-1)$-sphere,
    then an atlas of $\partial \mathcal{B}_{p_e}$ is obtained by  $\{\mathbf{b}_{i}( \bm{\theta}) \ | \ \bm{\theta} \in \Theta_{i} \}_{i = 1}^m$, where
    $\mathbf{b}_{i}( \bm{\theta})$ is obtained from $\mathbf{u}_{i}( \bm{\theta})$ using the following relation:
    \begin{eqnarray}
        \label{eq:EC_param_rep}
        \mathbf{b}(\bm{\theta}) =  C_{p_e}(\bm{\theta}) \mathbf{u}(\bm{\theta}) 
        + \nabla_{\bm{\theta}} \mathbf{u}(\bm{\theta}) g^{-1}(\bm \theta)(\nabla_{\bm{\theta}} C_{p_e}(\bm{\theta}))^{T}, 
    \end{eqnarray}
    and where $ g(\bm{\theta}) = \nabla_{\bm{\theta}} \mathbf{u}(\bm{\theta})^{T} \nabla_{\bm{\theta}} \mathbf{u}(\bm{\theta})$ is the metric 
    tensor of the $(n-1)$-sphere induced by the parametrization $\mathbf{u}(\bm{\theta})$.    
\end{theorem}

The proof of Theorem \ref{thm:param} is given in Appendix \ref{app:proof_thm_param}. 
Note that  Theorem \ref{thm:param} gives an analytic expression for the environmental contour (i.e. $\mathbf{b}_{i}(\bm{\theta})$) 
in terms of the $p_e$-level percentile function $C_{p_e}(\bm{\theta})$. 
Thus, given a specific parametrization and a differentiable approximation of $C_{p_e}(\bm{\theta})$ it is possible to compute $\mathbf{b}(\bm{\theta})$ directly, as an alternative 
to explicitly constructing a Voronoi cell as described in section \ref{sec:EC_as_VC}. 
One common parametrization in the $n$-dimensional case is given by $\mathbf{u}(\bm{\theta}) = (u_0,u_1,\dots,u_{n-1})$ with $u_i=\cos{\theta_i}\prod_{j=0}^{i-1}\sin{\theta_j}$ for $i=0,1,\dots,n-2$ and $u_{n-1}=\prod_{j=0}^{n-2}\sin{\theta_j}$, where  $\theta_i \in [0, \pi)$ for $i=1,2,\dots,n-2$ and $\theta_{n-2} \in [0, 2\pi)$.
The corresponding induced metric tensor has entries $g_{0,0}= 1$, $g_{i,i}=\prod_{j=0}^{i-1}\sin{\theta_j}^2$ for $i=0,1,\dots,n-2$ and $g_{i,j}=0$ if $i\neq j$.

It would be desirable to have a criterion for $ C_{p_e}(\bm{\theta})$ that guarantees that 
$\mathbf{b}_{i}(\bm{\theta})$ represent a proper environmental contour. 
To obtain such a criterion, we will need a couple of intermediate results given 
in the following to Lemmas. 
\begin{lemma}
    \label{lemma:existence}
    The random variable $\mathbf X$ admits a proper convex environmental contour
    with respect to $p_{e} \in (0, 0.5)$ if and only if the following holds:\\

    For any $\mathbf u' \in \mathcal{U}$, there exists some $\mathbf o \in \Pi_{p_e}(\mathbf u')$
    such that $C^{\mathbf o}_{p_e}(\mathbf u) \geq 0$ for all $\mathbf u \in \mathcal{U}$.
\end{lemma}

\begin{lemma}
    \label{lemma:b_theta_pi}
    Assume the percentile function $C_{p_e}(\bm{\theta})$ is twice differentiable and that
    $\mathbf{u}(\bm{\theta}): \Theta \rightarrow \mathcal{U}$ is regular 
    ($\nabla \mathbf{u}(\bm{\theta})$ exists and has full rank for all $\bm{\theta}$).
    Let $\mathbf{b}(\bm{\theta})$ be defined as in \eqref{eq:EC_param_rep}. 
    Then 
    \begin{equation*}
        \mathbf{u}(\bm \theta)^T \mathbf{b}(\bm{\theta}) = C_{p_e}(\bm{\theta})
        \text{ and }
        \mathbf{u}(\bm \theta)^T \nabla \mathbf{b}(\bm{\theta}) = \bm{0}
    \end{equation*}
    for all $\bm{\theta} \in \Theta$. This means that $\Pi_{p_e}(\bm{\theta})$ is tangential to 
    $\mathbf b (\Theta)$ at the point $\mathbf{b}(\bm{\theta})$.
\end{lemma}

Lemma \ref{lemma:existence} comes as a 
consequence of Lemma \ref{lemma:cont_vor}, and the proof is given in Appendix \ref{app:proof_lemma_existence}. In Appendix \ref{app:proof_lemma_b_theta_pi} 
we present the proof of Lemma \ref{lemma:b_theta_pi}, which states that 
for any $\bm{\theta}$, the hyperplane $\Pi_{p_e}(\bm{\theta})$ is tangential to 
$\mathbf b (\Theta)$ at the point $\mathbf{b}(\bm{\theta})$.

Armed with these results we can prove the following criteria for existence.

\begin{theorem}[Existence of proper convex environmental contours]
    \label{thm:existence}
    Let $\mathbf X$ be any $n$-dimensional random variable where the percentile function 
    $C_{p_e}(\cdot)$ is differentiable on the unit $(n-1)$-sphere.
    Let $\{\mathbf{u}_{i}( \bm{\theta}) \ | \ \bm{\theta} \in \Theta_{i} \}_{i = 1}^m$ be a regular parametrization of the unit $(n-1)$-sphere,
    and define for any $\mathbf{u}( \bm{\theta}) = \mathbf{u}_{i}( \bm{\theta})$ the function 
    \begin{equation}
        \kappa(\bm{\theta} | \bm{\theta}') = C_{p_e}^{\mathbf{b}(\bm{\theta}')}(\bm{\theta})
        = C_{p_e}(\bm{\theta}) - \mathbf{u}( \bm{\theta}) \cdot \mathbf{b}(\bm{\theta}'),
    \end{equation}
    where $C_{p_e}(\bm{\theta}) = C_{p_e}(\mathbf u (\bm{\theta}))$ and $\mathbf{b}(\bm{\theta}')$ 
    is given by \eqref{eq:EC_param_rep} with $\bm{\theta} = \bm{\theta}'$. 
    
    Then the following are equivalent:
    \begin{center}
    \begin{minipage}{0.8\textwidth}
        \begin{enumerate}
            \item $\mathbf X$ admits a proper convex environmental contour.
            \item The hypersurface given by the parametrization $b(\bm{\theta})$ in \eqref{eq:EC_param_rep} 
            is the boundary of a closed convex set.
            \item $
            \kappa(\bm{\theta} | \bm{\theta}')  \geq 0 \text{ for all } 
            \mathbf{u}( \bm{\theta}) = \mathbf{u}_{i}( \bm{\theta}), 
            \bm{\theta}, \bm{\theta}' \in \Theta_{i}, \text{ and } i = 1, \dots, m.$
            \item $\kappa(\bm{\theta} | \bm{\theta}')$ attains its global minimum at $\bm{\theta} = \bm{\theta}'\text{ for all } 
            \mathbf{u}( \bm{\theta}) = \mathbf{u}_{i}( \bm{\theta}), i = 1, \dots, m$.
        \end{enumerate}
    \end{minipage}
    \end{center}
\end{theorem}

The proof of Theorem \ref{thm:existence} is provided in Appendix \ref{app:proof_thm_existence}.
In the $2$-dimensional case with polar coordinates, one can also show that existence is equivalent to the criterion that either $C_{p_e}(\theta) + C_{p_e}''(\theta) > 0$ 
or $C_{p_e}(\theta) + C_{p_e}''(\theta) < 0$ for all $\theta \in [0, 2\pi)$
(see \cite{Huseby:stk4400}). As a consequence of Theorem \ref{thm:existence}, 
we can obtain the following similar result stated in Corollary \ref{cor:hess}.

\begin{corollary}
    \label{cor:hess}
    Assume the $n$-dimensional random variable $\mathbf X$ admits a proper convex environmental contour, 
    and that $C_{p_e}(\bm{\theta})$ is two times differentiable. 
    Then $Hess( C_{p_e}(\bm{\theta}))  + g(\bm \theta) C_{p_e}(\bm{\theta}) $ is positive semi-definite for all $\bm{\theta}\in \Theta$, where $Hess(\cdot)$ is the Hessian operator on the $(n-1)$-sphere and $g(\bm \theta)$ is the $(n-1)$-sphere metric tensor. 
\end{corollary}

The proof of Corollary \ref{cor:hess} is given in Appendix \ref{app:proof_cor_hess}. Note that the
metric tensor on the unit circle is simply $g=1$, so the $2$-dimensional
version of Corollary \ref{cor:hess} states that $C_{p_e}(\theta) + C_{p_e}''(\theta) \geq 0$.
As a stronger version of the statement holds in the $2$-dimensional case, we 
might conjecture that the criterion in Corollary \ref{cor:hess} with strict positive definiteness could hold 
as both a necessary and sufficient condition for existence, but we have currently not explored 
this further in any detail. 

\section{Practical application of the Voronoi method for environmental contour approximation}
\label{sec:algs}
In Section \ref{sec:EC_as_VC} we outlined a potential procedure for 
approximating environmental contours using the Voronoi-representation. 
Based on this idea, we present the steps involved in Algorithm \ref{alg:vanilla} below, 
followed up by a discussion on how each step may be implemented in practice. 
 
\begin{alg}
    Approximating $\mathcal B_{p_e}$ using the Voronoi method
    \begin{enumerate}
        \item Select a set of unit vectors $U = \{ \mathbf{u}_{j}\}^M_{j=1}$.
        \item Estimate $\widehat{C}_{p_{e}}(\mathbf{u}_{j}) \approx C_{p_{e}}(\mathbf{u}_{j})$ for each $j = 1, \dots , M$.       
        \item Compute $\widehat{\mathcal S}^{\mathbf o}_{p_e}(U)$, 
        using $\widehat{C}^{\mathbf o}_{p_{e}}(\mathbf{u}_{j})$ in \eqref{eq:S}, for some $\mathbf o \in \mathcal B_{p_e} \setminus \partial \mathcal B_{p_e}$.
        \item Compute the approximation $\widehat{\mathcal{B}}_{p_{e}} = \vor{\mathbf o, \widehat{\mathcal S}^{\mathbf o}_{p_e}(U)}$.
        \item Check that each point in $\widehat{\mathcal S}^{\mathbf o}_{p_e}(U)$ is connected with $\mathbf o$ 
        in the Delaunay triangulation of the point set 
        $\{ \mathbf o \} \cup S^{\mathbf o}_{p_e}(U)$. 
    \end{enumerate}
    \label{alg:vanilla}
\end{alg}

\textbf{Step 1}: The algorithm will produce finer approximations as more unit vectors are included. 
However, the main computational burden is usually related to the estimation of $C_{p_{e}}(\mathbf{u}_{j})$ for each 
unit vector, so the number of unit vectors is often decided by the desired run-time of the entire algorithm. 
In applications such as design of marine structures, there might be knowledge related to which directions that are the most informative, 
and the set $U$ might be chosen on this basis. Alternatively, a uniform selection may be applied. 
One way to generate uniform random samples from the unit $(n-1)$-sphere 
is to let $\mathbf{u}_{j} = \mathbf{v}_{j} / \norm{\mathbf{v}_{j}}$ where 
$\mathbf{v}_{j} = (v_{1, j}, \dots, v_{n, j})$ and all $v_{1, j}$
are i.i.d. Gaussian \citep{Marsaglia:1972:nsphere}. 

\textbf{Step 2}:
In practice, we might not be able to compute $C_{p_{e}}(\mathbf{u}_{j})$ exactly. However, this can be estimated based on a finite number of 
Monte Carlo samples from the joint distribution, in the same way as outlined in \cite{Vanem:EnvCont12, Vanem:EnvCont14}. 
The estimation error will depend on the sample size and may in principle be reduced to an acceptable level by increasing the number of samples, or for example using the importance sampling scheme proposed in \cite{HusebyVN:ESREL2014}. 
Moreover, if one were to apply conservative estimates, i.e. 
$\widehat{C}_{p_{e}}(\mathbf{u}_{j}) \geq C_{p_{e}}(\mathbf{u}_{j})$, 
this would produce a conservative (larger) environmental contour approximation as well.

\textbf{Step 3}:
In order to compute $\widehat{\mathcal S}^{\mathbf o}_{p_e}(U)$, 
we first need some point of reference $\mathbf o$ from the interior of $\mathcal B_{p_e}$.
The criterion that $C^{\mathbf o}_{p_e}(\mathbf u)>0$ for any $\mathbf u \in \mathcal U$ (see Lemma \ref{lemma:cont_vor}) 
can be used to identify if the selected origin $\mathbf o$ is not in the interior of $\mathcal B_{p_e}$.
We can then also observe that, in the case where we want to replace the origin $\mathbf{o}$ with some new point $\mathbf{o}^{*}$, 
the new set $\mathcal S^{\mathbf{o}^{*}}_{p_e}$ can be computed using that $C^{\mathbf{o}^{*}}_{p_e}(\mathbf u) = C^{\mathbf{o}}_{p_e}(\mathbf u) + \mathbf{u} \cdot (\mathbf{o} - \mathbf{o}^{*})$, and hence
\begin{equation}
    \mathbf s_{p_e}^{\mathbf{o}^{*},\mathbf u} = \mathbf s_{p_e}^{\mathbf{o},\mathbf u} + 2\mathbf{u} \cdot (\mathbf{o} - \mathbf{o}^{*}) \mathbf{u} - (\mathbf{o} - \mathbf{o}^{*}).
\end{equation}
This means that the estimates $\widehat{C}_{p_{e}}(\mathbf{u}_{j})$ can be reused, as going from 
$\widehat{\mathcal S}^{\mathbf o}_{p_e}(U)$ to $\widehat{\mathcal S}^{\mathbf{o}^{*}}_{p_e}(U)$ is a simple linear transformation. 
We may also note the geometric interpretation, by observing that the added term $2\mathbf{u} \cdot (\mathbf{o} - \mathbf{o}^{*}) \mathbf{u} - (\mathbf{o} - \mathbf{o}^{*})$ 
is the reflection of the point $(\mathbf{o} - \mathbf{o}^{*})$ with respect to the unit vector $\mathbf{u}$.
As both checking whether $C^{\mathbf o}_{p_e}(\mathbf u)>0$ and moving the origin $C^{\mathbf o}_{p_e}(\mathbf u) \rightarrow C^{\mathbf{o}^{*}}_{p_e}(\mathbf u)$
are cheap computationally, one could derive an iterative procedure to determine $\mathbf o$. 
Alternatively, finding the point $\mathbf o$ with maximal distance to all hyperplanes under the 
restriction that $C^{\mathbf o}_{p_e}(\mathbf{u}_{j})>0$, which is equivalent to 
$C_{p_e}(\mathbf{u}_{j}) > \mathbf{u}_{j} \cdot \mathbf o$, for each $j = 1, \dots, M$ can 
be solved by linear programming. 
In our implementation, 
the geometric median of a set of samples from the joint distribution of $\mathbf X$ (the ones used to estimate $C_{p_{e}}(\mathbf{u}_{j})$ in Step 2) 
was selected as the origin $\mathbf o$. This choice of $\mathbf o$
will with high probability lie inside $\mathcal B_{p_e}$ for any $p_e>0.5$, 
and in our experiments we did not find the need to iterate further beyond this initial guess.  

\textbf{Step 4}:
Some of the motivation for this paper comes from the fact that the Voronoi tessellation is a well studied object. 
As a result, a wide range of software and programming languages come with efficient procedures for computing Voronoi cells, 
including Python/Scipy, R, Wolfram Language/Mathematica, Matlab and Octave. 
Moreover, Voronoi algorithms work in arbitrary dimensions, which is what makes the proposed algorithm agnostic to the dimensionality of $\mathbf X$.

\textbf{Step 5}:
This check comes as a consequence of Proposition \ref{prop:delaunay_proper_EC} and Corollary \ref{cor:delaunay_proper_EC}.
There are two scenarios that may cause this check to fail. 1) When the selected probability distribution does not 
admit a proper convex environmental contour with respect to the chosen target probability, and 2) 
when the percentile function $C_{p_e}(\mathbf u)$ is estimated with error. 
In the case where the check fails due to noise in the estimates $\widehat{C}_{p_{e}}(\mathbf{u}_{j})$, 
we can make refinements based on the relevant unit vectors. 
For instance, if it is found that the point $\hat{\mathbf{s}}_{k} \in \widehat{\mathcal S}^{\mathbf o}_{p_e}(U)$ 
corresponding to unit vector $\mathbf{u}_{k}$ is not connected with $\mathbf o$, the estimates 
$\widehat{C}_{p_{e}}(\mathbf{u}_{j})$ can be refined for relevant indices $j$. The relevant indices here, besides $j = k$, 
are the ones corresponding to points $\hat{\mathbf{s}}_{j}$ affecting the Delaunay triangulation 
in the vicinity of $\hat{\mathbf{s}}_{k}$, which are the points connected with $\hat{\mathbf{s}}_{k}$ and the neighbouring Delaunay simplices.
With reference to the previous step, we also note that the task of obtaining the Delaunay triangulation usually "comes for free", in the sense that 
available algorithms used to obtain the Voronoi tessellation do this by computing the Delaunay triangulation and taking the dual. \\

The goal of this numerical procedure presented in Algorithm \ref{alg:vanilla} is to provide a good approximation in the 
case where a proper convex environmental contour exists. 
In the case where a proper convex environmental contour \emph{does not} exist, 
one might still be interested in finding a \emph{valid} convex environmental contour
that is "as small as possible". That is, a convex set where the exceedance probability 
of each supporting half-space is less than or equal to $p_{e}$ (where it cannot be 
equal to $p_{e}$ for all supporting half-spaces as no \emph{proper} convex environmental contour exists).
We will end this section with a modified version of the algorithm 
to accommodate this scenario.

The contour $\partial \mathcal B_{p_e}$ corresponding to the boundary of a Voronoi cell $Vor(\mathbf o,\mathcal S^{\mathbf o}_{p_e})= \bigcap_{\mathbf u \in U} \Pi_{p_e}^-(\mathbf u)$
is only a valid and proper environmental contour if $\partial \mathcal B_{p_e}\cap  \Pi_{p_e}^-(\mathbf u) \neq \emptyset$ $\forall \mathbf u \in U$. 
Otherwise, it is invalid. 
We may however use an \emph{invalid} Voronoi contour to create a \emph{valid improper} contour by the following algorithm:
\begin{alg}
Let V be a Voronoi contour computed by Algorithm \ref{alg:vanilla} based on a set of unit vectors $U$.
\begin{enumerate} 
\item Initialise $Z=V$. 
\item For each direction $\mathbf u \in U$:
\begin{enumerate} 
\item Find the point $\mathbf v' \in V$ that is furthest out in direction $\mathbf u$, i.e.  $\mathbf v' = \underset{\mathbf v \in V}{\mathrm{argmax}} \{ \mathbf v \cdot \mathbf u\}$.
\item Compute the projection of $\mathbf v'$ onto the plane $ \Pi_{p_e}(\mathbf u)$, i.e. $\mathbf z = \mathbf v' + (C_{p_e}(\mathbf u) - \mathbf v' \cdot \mathbf u)\mathbf u $.
\item Update $Z \rightarrow Z \cup \{ \mathbf z \}$.
\end{enumerate} 
\item Compute the convex hull of $Z$. This is the corrected Voronoi contour.
\end{enumerate} 
\label{alg:corrected}
\end{alg}

\begin{figure}[h!]
    \center{\includegraphics[width=0.4\textwidth]{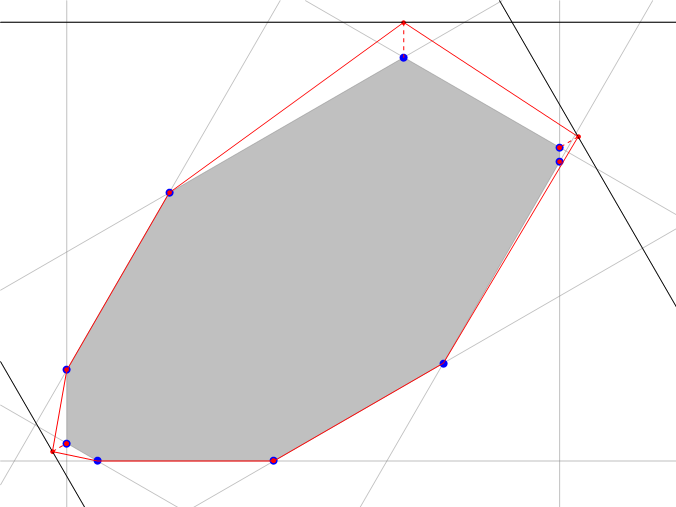}}
    \caption{Illustration of algorithm \ref{alg:corrected} to construct a valid environmental contour (red).}
	\label{fig:corrected_voronoi_method}
\end{figure}

The algorithm above guarantees a valid environmental contour with respect to $U$, because it intersect all the hyperplanes $\Pi_{p_e}^-(\mathbf u)$ $\forall \mathbf u \in U$ by construction.
The projection algorithm is illustrated in figure \ref{fig:corrected_voronoi_method}.

\begin{figure}[h]
    \center{\includegraphics[width=0.9\textwidth]{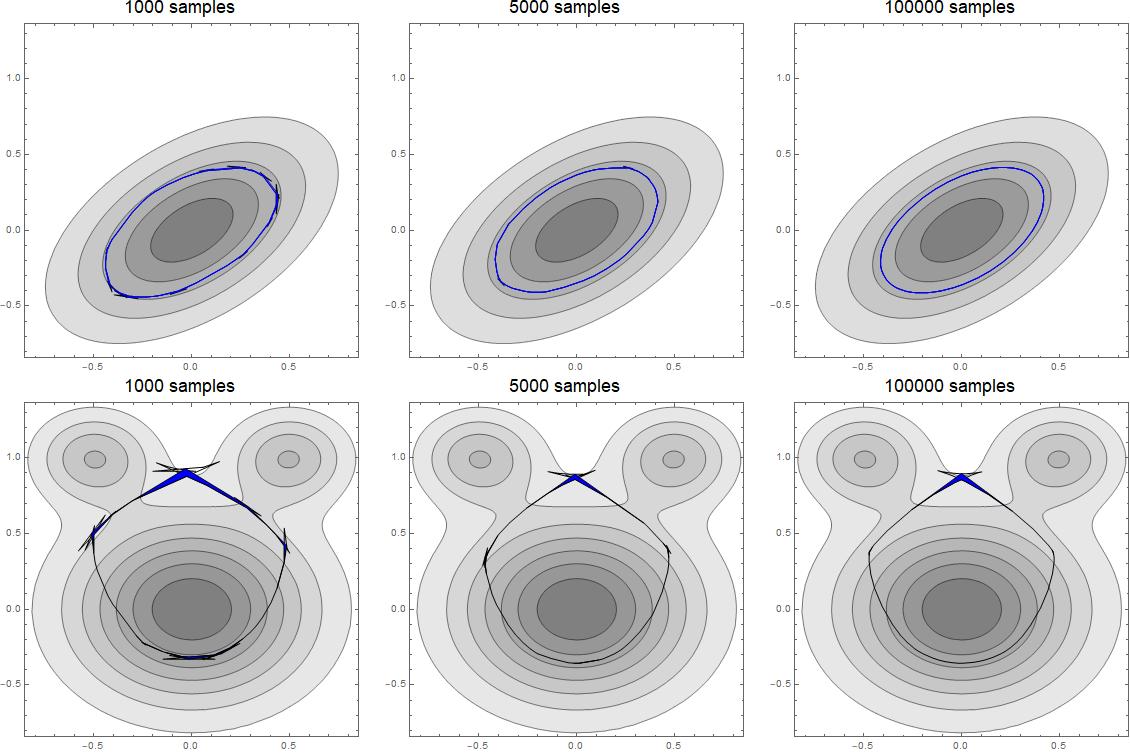}}
  \caption{
Top: Contours for a multinormal distribution, constructed using the direct method of \cite{Vanem:EnvCont12}. The loops disappear as the number of samples increased, indicating that the loops is a sampling issue. 
Bottom: Contours for a multimodal distribution constructed using the direct method of \cite{Vanem:EnvCont12}. The top loops does not disappear as the number of samples increase, indicating that the loops is a feature of the underlying distribution (i.e. the distribution does not admit a proper convex contour for the selected target probability).}
   \label{fig:micky_mouse}
\end{figure}

Figure \ref{fig:micky_mouse} shows two examples using the above algorithms and also the direct method 
presented in \cite{Vanem:EnvCont12}. First, a scenario where a proper convex environmental contour exists, 
and then a scenario where a proper environmental contour does not exist. 
The top row corresponds to a centered bivariate normal distribution with covariance $0.16\cdot [1 \  0.5;0.5 \ 1]$, and the bottom row
represents a Gaussian mixture;
$\mathbf{X} = 0.8\mathbf{X}_{1} + 0.1\mathbf{X}_{2} + 0.1\mathbf{X}_{3}$ where 
$\mathbf{X}_{1} \sim \mathcal{N}([0 \ 0]^{T}, 0.16I)$, $\mathbf{X}_{2} \sim \mathcal{N}([0.5 \ 1]^{T}, 0.04I)$
and $\mathbf{X}_{3} \sim \mathcal{N}([-0.5 \ 1]^{T}, 0.04I)$. The contours are computed with $p_{e} = 0.15$.

\section{Examples}

\subsection{2D example}

To illustrate the Voronoi approach in two dimensions, we use the same example as \cite{Vanem:EnvCont12}.
The environmental variables of interest are the significant wave height, $H_S$, and the zero-upcrossing wave period, $T_Z$. Their joint distribution is modelled using a conditional modelling approach \cite{ElzbietaBook12, EBG:JointDesc2015}, and can be expressed as 
\begin{align}
f_{H, T} ( h, t) = f_H(h) f_T(t | h).
\end{align}
Here, $f_H(h)$ is a 3-parameter Weibull distribution for significant wave height, with scale parameter $\alpha$, shape parameter $\beta$, and location parameter $\gamma$.
$f_T(t | h)$ is a conditional log-normal distribution for wave period, where the model parameters are functions of significant wave height, as outlined in e.g. \cite{DNVGLRP-C205, VanemB-G:JOMAE15}, i.e.
\begin{align} 
\begin{split}
\mu_T(h) = E\left(\ln T_Z | H_S = h \right) = a_1 + a_2 h^{a_3} \\
\sigma_T(h) = sd \left(\ln T_Z | H_S = h \right) = b_1 + b_2 e^{b_3 h}.
\end{split}
\end{align}
The parameter values used are listed in Table \ref{tab:2dparameters}.

\begin{table}[!ht]
	\centering \footnotesize
	\caption{Parameters assumed for the bivariate distribution of $H_S$ and $T_S$.}
	\begin{tabular}{ l r c c c }
		\toprule 
		\multicolumn{2}{l}{3-p Weibull ($H_S$)} & $\alpha$& $\beta$ & $\gamma$ \\
		\hline\noalign{\smallskip}
		 & & 2.776 & 1.471 & 0.8888 \\
		\toprule 
		\multicolumn{2}{l}{Conditional log-normal ($T_Z$)} & i = 1 & i = 2 & i = 3  \\
		\hline\noalign{\smallskip}
		\multirow{2}{*}{} & a$_i$ & 0.1000 & 1.4890 & 0.1901 \\
		& b$_i$ & 0.0400 & 0.1748 & -0.2243 \\				 
		\bottomrule
	\end{tabular}
	\label{tab:2dparameters}
\end{table}

\begin{figure}[h!]
    \center{\includegraphics[width=0.9\textwidth]{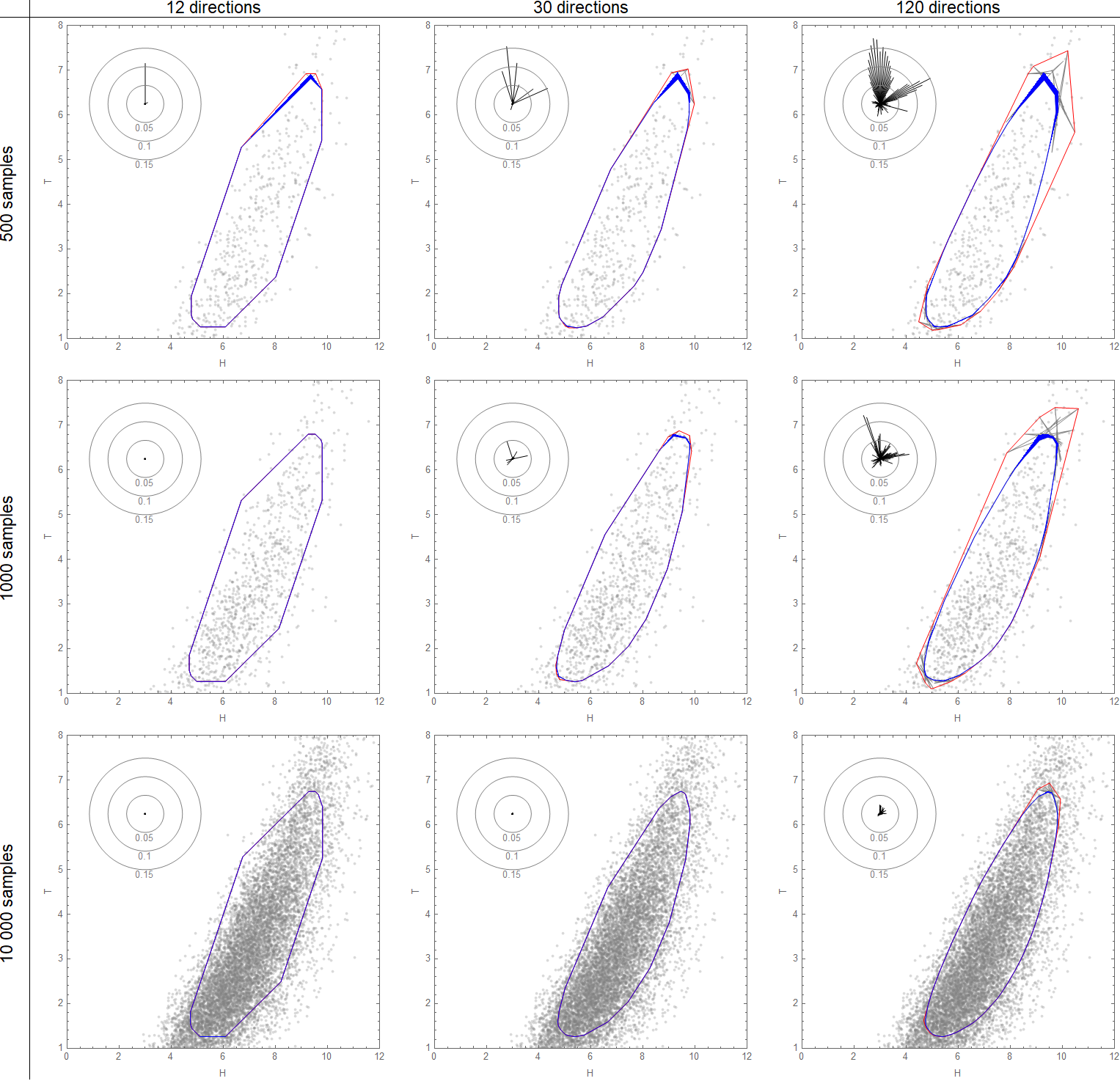}}
    \caption{Comparison of results, for $p_e=0.05$. The samples that contours are computed from are shown in grey. The black curves represent the direct sampling method of \cite{Vanem:EnvCont12}. The red curves represent the convex hull of the black curves. The blue regions represent the Voronoi methods; the inner boundary correspond to the simple Voronoi method, and the outer boundary correspond to the corrected Voronoi method. The insets show the error in different directions, i.e. the difference between the simple and corrected Voronoi methods.}
\label{fig:2dexample}
\end{figure}

Figure \ref{fig:2dexample} shows comparisons of results for different methods.
The number of samples that the contours are based on is varied in the rows, but the samples are identical within each row. The number of unit vectors used to compute the contours is varied in the columns. 

The direct sampling method of \cite{Vanem:EnvCont12} is drawn in black. This method does not guarantee convex contours, but sometimes produce loops. Keeping the samples fixed, the loops tend to be larger as the number of unit vectors increase, which is undesirable. However, the loops tend to get smaller with increased number of samples.
The convex hull of the black contours are drawn in red. Note that for the same number of samples, these red contours tend to get larger when the number of directions is increased, due to the larger loops. 

Contours based on the Voronoi method are shown in blue. More precisely, blue regions are plotted, where the inner boundary correspond to the simple Voronoi method (i.e. Algorithm \ref{alg:vanilla}), and the outer boundary correspond to the corrected Voronoi method (i.e. Algorithm \ref{alg:corrected}). 
Note that, unlike the other methods, the contours produced by the Voronoi methods do not diverge as the number of directions is increased. We also see that the shaded region is generally thin, indicating that the simple Voronoi method is a good approximation to the 'true' environmental contour.
The inset shows the error, i.e. the difference between the two Voronoi methods in the various directions. The directions with high error corresponds to directions where the direct method of  \cite{Vanem:EnvCont12} produces loops, i.e. the Voronoi method provides a warning for directions where more sampling may be needed. 

\subsection{3D example}
To illustrate the Voronoi approach in three dimensions, we include an example from \cite{Vanem3Dcontour17}. The environmental variables of interest are the significant wave height, $H_S$, the zero-upcrossing wave period, $T_Z$, and the 10-minute mean wind speed at a particular height, $U_{10}$. Their joint distribution is modelled using a conditional modelling approach \cite{ElzbietaBook12, EBG:JointDesc2015}, and can be expressed as 
\begin{align}
f_{H, T, U} ( h, t, u) = f_H(h) f_T(t | h) f_U(u | h).
\end{align}

$f_H(h)$ is a 3-parameter Weibull distribution for significant wave height, with scale parameter $\alpha$, shape parameter $\beta$, and location parameter $\gamma$.

$f_T(t | h)$ is a conditional log-normal distribution for wave period, where the model parameters are a function of significant wave height as outlined in e.g. \cite{DNVGLRP-C205, VanemB-G:JOMAE15}, i.e.
\begin{align} 
\begin{split}
\mu_T(h) = E\left(\ln T_Z | H_S = h \right) = a_1 + a_2 h^{a_3} \\
\sigma_T(h) = sd \left(\ln T_Z | H_S = h \right) = b_1 + b_2 e^{b_3 h}.
\end{split}
\end{align}
The parameters $a_i, b_i$, $i = 1, 2, 3$ are estimated from data. 

$f_U(u | h)$ is a conditional 2-parameter Weibull distribution with parameters modelled as functions of significant wave height as suggested by \cite{DNVGLRP-C205, BGH:JointEnvModRel91, BGH:JointEnvParam89}. The scale parameter,  $\lambda_U$, and shape parameter, $\kappa_U$, are modelled as
\begin{align}
\begin{split}
\lambda_U(h) = c_1 + c_2 h^{c_3} \\
\kappa_U(h) = d_1 + d_2 h^{d_3}.
\end{split}
\end{align}

For the significant wave height and wave period, parameters corresponding to average world wide operations of ships according to appendix C of \cite{DNVGLRP-C205} are assumed, as summarised in Table \ref{table:WeibullPara205}. For the conditional distribution of wind speed, the average sectoral parameters reported in \cite{BGH:JointEnvModRel91, BGH:JointEnvParam89} will be assumed, as summarised in Table \ref{table:WeibullPara205}. It is noted that the parameter $d_3$ is omitted in \cite{BGH:JointEnvModRel91}, so this is simply set to 1 in this study. 

\begin{figure}[h!]
    \center{\includegraphics[width=0.9\textwidth]{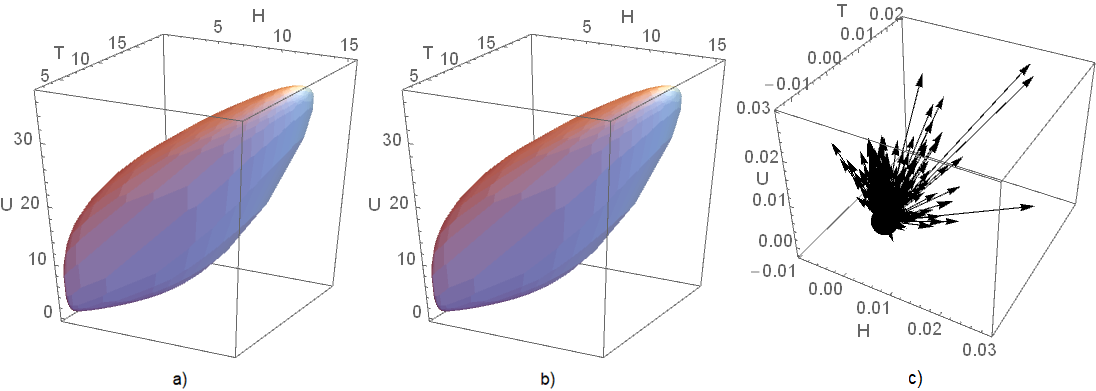}}
    \caption{
a) Approximate (invalid) environmental contour for 3D example, computed using the simple Voronoi method (i.e. Algorithm \ref{alg:vanilla}).
b) Valid (improper) environmental contour for 3D example, computed using the corrected Voronoi method (i.e. Algorithm \ref{alg:corrected}).
c) Difference between the corrected and simple Voronoi methods, showing that the simple method gives good approximation to a valid environmental contour. 
 }
\label{fig:3d_example}
\end{figure}

Figure \ref{fig:3d_example} shows the result of applying the Voronoi methods (simple and corrected) to the example described above. As can be seen, the simple method and corrected method are very similar, indicating that the simple Voronoi method is a good approximation for the 'true' environmental contour.

\begin{table}[!ht]
	\centering \footnotesize
	\caption{Parameters assumed for the trivariate distribution of $H_S$, $T_S$ and $U_{10}$.}
	\begin{tabular}{ l r c c c }
		\toprule 
		\multicolumn{2}{l}{3-p Weibull ($H_S$)} & $\alpha$& $\beta$ & $\gamma$ \\
		\hline\noalign{\smallskip}
		average World wide trade & & 1.798 & 1.214 & 0.856 \\
		\toprule 
		\multicolumn{2}{l}{Conditional log-normal ($T_Z$)} & i = 1 & i = 2 & i = 3  \\
		\hline\noalign{\smallskip}
		\multirow{2}{*}{average World wide trade} & a$_i$ & -1.010 & 2.847 & 0.075 \\
		& b$_i$ & 0.161 & 0.146 & -0.683 \\	
		\toprule 		
		\multicolumn{2}{l}{Conditional 2-p Weibull ($U_{10}$)} & i = 1 & i = 2 & i = 3  \\
		\hline\noalign{\smallskip}		
		\multirow{2}{*}{average directional sector} & c$_i$ & 2.58 & 0.12 & 1.60 \\	
		& d$_i$ & 4.6 & 2.05 & 1 \\					 
		\bottomrule
	\end{tabular}
	\label{table:WeibullPara205}
\end{table}

\section{Concluding remarks}
In this paper, a novel algorithm for constructing environmental contours has been presented, based on a geometric interpretation of environmental contours as Voronoi cells. One advantage of this approach is that many software libraries exist for Voronoi cell computation, making the algorithm simple to implement.  Another advantage is that the Voronoi method also makes it easy to compute environmental contours in higher than two dimensions. The Voronoi environmental contours are not guaranteed to be proper, but with a simple modification to the algorithm, valid environmental contours can always be constructed from improper Voronoi environmental contours.

The Voronoi geometric interpretation also has given new intuition and theoretical insights about environmental contours, including representation and existence theorems for proper convex environmental contours. The presented analytical formula provides another alternative algorithm to compute environmental contours. Interestingly, this formula has an analogy in shadow systems and can be interpreted as an inverse Gauss map \cite{Shephard:1964shadow,Epstein:shadow,Martini:2019bodies}. Further exploration of this correspondence between environmental contours and shadow functions could potentially reveal new insights in both domains, and potentially provide some information on the class of random variables for which proper environmental contours exist. 

\section*{Acknowledgements}
This work has been supported by grant 276282 from the Research Council of Norway (RCN)
and DNV GL Group Technology and Research.
Parts of the work has also been carried out within the research project ECSADES, with support from RCN under the MARTEC II ERA-NET initiative; project no. 249261/O80.   

\begin{appendices}
    \section{Proof of Lemma \ref{lemma:cont_vor}}
    \label{app:proof_lemma_cont_vor}
    Proving the first statement is trivial, as $\mathbf x \in \mathcal B_{p_e}$ by definition means that 
    $\mathbf u \cdot (\mathbf x - \mathbf o) \leq C^{\mathbf o}_{p_e}(\mathbf u)$ for any $\mathbf u \in \mathcal{U}$.
    So, in particular, we have that
    $\mathbf o \in \mathcal B_{p_e} \Leftrightarrow 0 = \mathbf u \cdot (\mathbf o - \mathbf o) \leq C^{\mathbf o}_{p_e}(\mathbf u)$.\\

    \noindent To prove the second statement we use that
    \begin{equation*}
        \mathbf x \in \mathcal B_{p_e} \setminus \partial \mathcal B_{p_e} 
        \Rightarrow \mathbf x \in \bigcap_{\mathbf u \in \mathcal{U}} \left( \Pi_{p_e}^-(\mathbf u) \setminus \partial \Pi_{p_e}^-(\mathbf u) \right).
    \end{equation*}
    That is, a point $\mathbf x$ in the interior of $\mathcal B_{p_e}$ is also in the 
    intersection of all interior half-spaces. 
    Hence, $\mathbf x \in \{ \mathbf x : \mathbf u \cdot (\mathbf x - \mathbf o ) < C^{\mathbf o}_{p_e}(\mathbf u) \}$
    for all $\mathbf u \in \mathcal{U}$. And so by the same argument as above we have that
    $\mathbf o \in \mathcal B_{p_e} \setminus \partial \mathcal B_{p_e} \Rightarrow
    0 = \mathbf u \cdot (\mathbf o - \mathbf o) < C^{\mathbf o}_{p_e}(\mathbf u) \ \forall \mathbf u \in \mathcal{U}$.
    
    To prove the converse, we first observe that if $\mathbf o \in \partial \mathcal B_{p_e}$, then 
    then there exists some $\mathbf u^* \in \mathcal{U}$ where $\mathbf o \in \partial \Pi_{p_e}^-(\mathbf u^*)$ (by the supporting hyperplane theorem) 
    which means that $C^{\mathbf o}_{p_e}(\mathbf u^*) = 0$, and if $\mathbf o \notin \mathcal B_{p_e}$ then we have already 
    shown that $C^{\mathbf o}_{p_e}(\mathbf u^*) < 0$ for some $\mathbf u^*$.
    Putting this together we get that $\mathbf o \notin \mathcal B_{p_e} \setminus \mathcal B_{p_e}
    \Rightarrow \exists \mathbf u^* \in \mathcal{U} \text{ s.t. } C^{\mathbf o}_{p_e}(\mathbf u^*) \leq 0$,
    and hence $C^{\mathbf o}_{p_e}(\mathbf u) > 0 \ \forall \mathbf u \in \mathcal{U}   
    \Rightarrow \mathbf o \in \mathcal B_{p_e} \setminus \mathcal B_{p_e}$.\\
    
    \noindent As for the final statement, we first recall that a point $\mathbf x$ is in $\vor{\mathbf o, \mathcal S^{\mathbf o}_{p_e}(U)}$
    if and only if $\norm{\mathbf x - \mathbf o} \leq \norm{\mathbf x - \mathbf s_{p_e}^{\mathbf o,\mathbf u}}$,
    or equivalently $\norm{\mathbf x - \mathbf o}^{2} \leq \norm{\mathbf x - \mathbf s_{p_e}^{\mathbf o,\mathbf u}}^{2}$,
    for any $\mathbf u \in U$.
    We first observe that 
    \begin{equation}
        \label{eq:proof_norm_exp}
        \norm{\mathbf x - \mathbf s_{p_e}^{\mathbf o,\mathbf u}}^{2}
        = \norm{\mathbf x - \mathbf o - 2C^{\mathbf o}_{p_e}(\mathbf u) \mathbf u}^{2}
        = \norm{\mathbf x - \mathbf o}^{2} + 4(C^{\mathbf o}_{p_e}(\mathbf u))^{2} - 4C^{\mathbf o}_{p_e}(\mathbf u)(\mathbf x - \mathbf o)\cdot \mathbf u,
    \end{equation}
    and so,
    \begin{equation*}
        \norm{\mathbf x - \mathbf o}^{2} \leq \norm{\mathbf x - \mathbf s_{p_e}^{\mathbf o,\mathbf u}}^{2} 
        \Leftrightarrow C^{\mathbf o}_{p_e}(\mathbf u)(\mathbf x - \mathbf o)\cdot \mathbf u \leq (C^{\mathbf o}_{p_e}(\mathbf u))^{2}.
    \end{equation*}
    Hence, using the second statement of the Lemma, we have that if $\mathbf o \in \mathcal B_{p_e} \setminus \partial \mathcal B_{p_e}$
    then $C^{\mathbf o}_{p_e}(\mathbf u) > 0$, and so 
    $\norm{\mathbf x - \mathbf o}^{2} \leq \norm{\mathbf x - \mathbf s_{p_e}^{\mathbf o,\mathbf u}}^{2} 
    \Leftrightarrow (\mathbf x - \mathbf o)\cdot \mathbf u \leq C^{\mathbf o}_{p_e}(\mathbf u)$
    for any $\mathbf u \in U$ which completes the proof.

    \qed

    \section{Proof of Proposition \ref{prop:main}}
    \label{app:proof_prop_main}
    
    First we recall that by definition $\mathcal B_{p_e} = \bigcap_{\mathbf u \in \mathcal U} \Pi_{p_e}^-(\mathbf u)$. 
    Using Lemma \ref{lemma:cont_vor} we then have
    $\mathcal B_{p_e} = \vor{\mathbf o, \mathcal S^{\mathbf o}_{p_e}(\mathcal{U})}$, and also 
    $\vor{\mathbf o, \mathcal S^{\mathbf o}_{p_e}(U_{i})} = \bigcap_{\mathbf u \in U_{i}} \Pi_{p_e}^-(\mathbf u)$
    for $i = 1, 2$. \\

    \noindent Since $U_{1} \subseteq U_{2} \subseteq \mathcal{U}$ the proof is completed by observing that

    \begin{equation*}
        \bigcap_{\mathbf u \in \mathcal U} \Pi_{p_e}^-(\mathbf u)
        \subseteq \bigcap_{\mathbf u \in U_{2}} \Pi_{p_e}^-(\mathbf u)
        \subseteq \bigcap_{\mathbf u \in U_{1}} \Pi_{p_e}^-(\mathbf u).
    \end{equation*}

    \qed

    \section{Proof of Proposition \ref{prop:delaunay_proper_EC}}
    \label{app:proof_prop_delaunay}
    The proof will follow from the Voronoi-Delaunay duality, which 
    tell us that the Voronoi cells 
    are convex polytopes with vertices corresponding to circumcenters of the Delaunay simplices. 
    In particular, the vertices of $\vor{\mathbf o, \mathcal S^{\mathbf o}_{p_e}(U)}$ 
    are the circumcenters of the simplices in
    $\{ \tau \in \mathcal{D} \ | \ \mathbf o \in \tau \}$, where $\mathcal{D}$ is any 
    Delaunay triangulation of the point set $\{ \mathbf o \} \cup S^{\mathbf o}_{p_e}(U)$.

    Assume that $\mathcal{D}$ is such a Delaunay triangulation, and that there exists a point $\mathbf{s}^{*} \in \mathcal S^{\mathbf o}_{p_e}(U)$ 
    such that $\mathbf{s}^{*}$ and $\mathbf o$ are not connected by $\mathcal{D}$. 
    This means (by definition) that any simplex in $\mathcal{D}$ containing $\mathbf o$ does not contain $\mathbf{s}^{*}$, and vice versa. 
    Hence, 
    \begin{equation*}
        \vor{\mathbf o, \mathcal S^{\mathbf o}_{p_e}(U)} = \vor{\mathbf o, \mathcal S^{\mathbf o}_{p_e}(U) \setminus \{ \mathbf{s}^{*} \}}.
    \end{equation*}
    We now let $\mathbf{u}^{*} \in U$ denote the unit vector corresponding to $\mathbf{s}^{*}$, 
    i.e. $\mathbf{s}^{*} = \mathbf s_{p_e}^{\mathbf o,\mathbf{u}^{*}}$.
    Making use of Lemma \ref{lemma:cont_vor} we then observe that 
    \begin{equation}
        \label{eq:app_vert1}
        \vor{\mathbf o, \mathcal S^{\mathbf o}_{p_e}(U)} = \vor{\mathbf o, \mathcal S^{\mathbf o}_{p_e}(U) \setminus \{ \mathbf{s}^{*} \}}
        \Rightarrow
        \bigcap_{\mathbf u \in U} \Pi_{p_e}^-(\mathbf u) = \bigcap_{\mathbf u \in U \setminus \{ \mathbf{u}^{*} \}} \Pi_{p_e}^-(\mathbf u).
    \end{equation}
    This means that, either 1) $\Pi_{p_e}(\mathbf{u}^{*}) \cap \vor{\mathbf o, \mathcal S^{\mathbf o}_{p_e}(U)} = \emptyset$,
    or 2) that there exists some vertex $\mathbf{v}^{*}$ of $\vor{\mathbf o, \mathcal S^{\mathbf o}_{p_e}(U)}$ 
    such that $\mathbf{v}^{*} \in \Pi_{p_e}(\mathbf{u}^{*}) \cap \vor{\mathbf o, \mathcal S^{\mathbf o}_{p_e}(U)}$.
    From Proposition \ref{prop:main} we have that 
    $\mathcal B_{p_e} \subseteq \vor{\mathbf o, \mathcal S^{\mathbf o}_{p_e}(U)}$. Since we assume that 
    $\partial \mathcal B_{p_e}$ is a proper convex environmental contour, $\Pi_{p_e}(\mathbf{u}^{*}) \cap \mathcal B_{p_e} \neq \emptyset$, and so 
    \begin{equation}
        \label{eq:app_vert2}
        \Pi_{p_e}(\mathbf{u}^{*}) \cap \vor{\mathbf o, \mathcal S^{\mathbf o}_{p_e}(U)} \neq \emptyset.
    \end{equation}
    From \eqref{eq:app_vert1} and \eqref{eq:app_vert2} we can therefore conclude that 
    there exists some vertex $\mathbf{v}^{*}$ of $\vor{\mathbf o, \mathcal S^{\mathbf o}_{p_e}(U)}$ 
    such that $\mathbf{v}^{*} \in \Pi_{p_e}(\mathbf{u}^{*}) \cap \vor{\mathbf o, \mathcal S^{\mathbf o}_{p_e}(U)}$.

    We then observe that
    \begin{equation}
        \label{eq:sphere_v_star}
        \mathbf{v}^{*} \in \Pi_{p_e}(\mathbf{u}^{*}) \Rightarrow \norm{\mathbf{v}^{*} - \mathbf o} = \norm{\mathbf{s}^{*} - \mathbf o}.
    \end{equation}
    This follows from the definition of $\Pi_{p_e}(\cdot)$ and the set $S^{\mathbf o}_{p_e}(U)$,
    which says that $\mathbf{s}^{*}$ is the reflection of $\mathbf o$ with respect to the hyperplane $\Pi_{p_e}(\mathbf{u}^{*})$.
    Now, since $\mathbf{v}^{*}$ is also a vertex of $\vor{\mathbf o, \mathcal S^{\mathbf o}_{p_e}(U)}$,
    then $\mathbf{v}^{*}$ is the circumcenter of a Delaunay simplex 
    $\tau$, with $\mathbf o \in \tau$. From \eqref{eq:sphere_v_star} we see that $\mathbf{s}^{*}$ 
    also lies on this circum-hypersphere, together with $\mathbf o$.
    Hence, if the Delaunay triangulation $\mathcal{D}$ was unique, 
    we could conclude that $\{ \mathbf{s}^{*}, \mathbf o \} \subset \tau \in \mathcal{D}$, which contradicts 
    the initial assumption that $\mathbf{s}^{*}$ and $\mathbf o$ are not connected in $\mathcal{D}$.
    
    In the case where there is no \emph{unique} Delaunay triangulation of the point set 
    $\{ \mathbf o \} \cup S^{\mathbf o}_{p_e}(U)$, the fact that 
    $\mathbf{s}^{*}$ and $\mathbf o$ lie on the same circum-hypersphere of \emph{some} 
    Delaunay simplex $\tau$ lets us conclude that there exists \emph{some} Delaunay 
    triangulation $\mathcal{D}'$
    where $\mathbf{s}^{*}$ and $\mathbf o$ are part of the same simplex. 
    We can therefore conclude that, if there exists a Delaunay triangulation $\mathcal{D}$ that does not connect 
    $\mathbf{s}^{*}$ and $\mathbf o$, then there must exist a different Delaunay triangulation $\mathcal{D}'$ that 
    connects $\mathbf{s}^{*}$ and $\mathbf o$.
    
    \qed

    \section{Proof of Proposition \ref{prop:sphere_property}}
    \label{app:proof_prop_sphere_property}
    For any $\mathbf{u} \in \mathcal{U}$ we first recall that the existence of some $\mathbf b \in \Pi_{p_e}(\mathbf{u}) \cap \partial \mathcal{B}_{p_e}$
    follows from the definition of proper convex environmental contours. 
    We then note that, as any element of $\mathcal S^{\mathbf o}_{p_e}(\mathcal{U})$ is of the form $\mathbf s_{p_e}^{\mathbf o,\mathbf{u}} =\mathbf o + 2C^{\mathbf o}_{p_e}(\mathbf{u}) \mathbf{u}$, 
    we have that
    \begin{equation}
        \begin{split}
            \norm{\mathbf s_{p_e}^{\mathbf o,\mathbf{u}} - \mathbf b}^{2}
            &= \norm{\mathbf o - \mathbf b + 2C^{\mathbf o}_{p_e}(\mathbf u) \mathbf u}^{2}
            = \norm{\mathbf o - \mathbf b}^{2} + 4(C^{\mathbf o}_{p_e}(\mathbf u))^{2} + 4C^{\mathbf o}_{p_e}(\mathbf u)(\mathbf o - \mathbf b)\cdot \mathbf u. \\
            & = \norm{\mathbf o - \mathbf b}^{2} + 4C^{\mathbf o}_{p_e}(\mathbf u)\left( C^{\mathbf o}_{p_e}(\mathbf u) - (\mathbf b - \mathbf o)\cdot \mathbf u \right).
        \end{split} 
        \label{eq:prop_sphere_proof}   
    \end{equation}
    Now if $\mathbf b \in \Pi_{p_e}(\mathbf{u})$ we have that $(\mathbf b - \mathbf o)\cdot \mathbf u = C^{\mathbf o}_{p_e}(\mathbf u)$ (by definition), 
    and hence $\norm{\mathbf s_{p_e}^{\mathbf o,\mathbf{u}} - \mathbf b}^{2} = \norm{\mathbf o - \mathbf b}^{2}$, which means 
    that $\mathbf s_{p_e}^{\mathbf o,\mathbf{u}} \in \partial \mathcal{W}^{\mathbf o}(\mathbf b)$.

    The statement that $\mathcal S^{\mathbf o}_{p_e}(\mathcal{U}) \cap \mathcal{W}^{\mathbf o}(\mathbf b) \subseteq \partial \mathcal{W}^{\mathbf o}(\mathbf b)$
    means that there are no $\mathbf{u}' \in \mathcal{U}$ such that $\mathbf s_{p_e}^{\mathbf o,\mathbf{u}'}$ lies in the interior of the ball $\mathcal{W}^{\mathbf o}(\mathbf b)$. 
    Assume, on the contrary, that there exists some $\mathbf s_{p_e}^{\mathbf o,\mathbf{u}'} \in \mathcal{W}^{\mathbf o}(\mathbf b) \setminus \partial \mathcal{W}^{\mathbf o}(\mathbf b)$.
    Then $\norm{\mathbf s_{p_e}^{\mathbf o,\mathbf{u}'} - \mathbf b} < \norm{\mathbf o - \mathbf b}$ by definition. 
    From \eqref{eq:prop_sphere_proof} we then have that 
    $4C^{\mathbf o}_{p_e}(\mathbf{u}')\left( C^{\mathbf o}_{p_e}(\mathbf{u}') - (\mathbf b - \mathbf o)\cdot \mathbf{u}' \right) < 0$.
    We have assumed that $\mathbf o \in \mathcal B_{p_e} \setminus \partial \mathcal B_{p_e}$, and so by Lemma \ref{lemma:cont_vor} 
    $C^{\mathbf o}_{p_e}(\mathbf{u}') > 0$. Hence, 
    \begin{align*}
        \mathbf s_{p_e}^{\mathbf o,\mathbf{u}'} \in \mathcal{W}^{\mathbf o}(\mathbf b) \setminus \partial \mathcal{W}^{\mathbf o}(\mathbf b) 
        \Rightarrow C^{\mathbf o}_{p_e}(\mathbf{u}') - (\mathbf b - \mathbf o)\cdot \mathbf{u}' < 0.
    \end{align*}
    But this means that $\mathbf b \in \Pi^{+}_{p_e} (\mathbf{u}')$, which is impossible when $\mathbf b \in \partial \mathcal{B}_{p_e}$.

    \qed

	\section{Proof of Lemma \ref{lemma:au}}
        \label{app:proof_lemma_au}

        We first observe that the condition 1) is just a different way of stating that 
        a point is on the hyperplane $\Pi_{p_e}(\mathbf u)$ 
        (alternatively, compute the norms as in \eqref{eq:prop_sphere_proof} and note that $C^{\mathbf o}_{p_e}(\mathbf u) > 0$).        
        That is, for any $\mathbf x \in \mathbb{R}^{n}$, 
        we have $\mathbf x \in \Pi_{p_e}(\mathbf u) \Leftrightarrow \norm{\mathbf{x} - \mathbf{o}} = \norm{\mathbf s_{p_e}^{\mathbf o,\mathbf u} - \mathbf{x}}$.

        Hence, $\mathbf a \in \Pi_{p_e}(\mathbf u)$ by condition 1). Then, by Proposition \ref{prop:sphere_property}
        there exists some $\mathbf b \in \Pi_{p_e}(\mathbf{u}) \cap \partial \mathcal{B}_{p_e}$
        where $\mathcal S^{\mathbf o}_{p_e}(\mathcal{U}) \cap \mathcal{W}^{\mathbf o}(\mathbf b) \subseteq \partial \mathcal{W}^{\mathbf o}(\mathbf b)$, 
        and $\mathbf s_{p_e}^{\mathbf o,\mathbf{u}} \in \partial \mathcal{W}^{\mathbf o}(\mathbf b)$.
        This means that the $n$-dimensional closed ball $\mathcal{W}^{\mathbf o}(\mathbf b)$,
        centered at $\mathbf b$ with radius $\norm{\mathbf b - \mathbf o}$ is tangent to 
        $\mathcal S^{\mathbf o}_{p_e}(\mathcal{U})$ at the point $\mathbf s_{p_e}^{\mathbf o,\mathbf{u}}$. 
        As both $\mathcal S^{\mathbf o}_{p_e}(\mathcal{U})$ and $\mathcal{W}^{\mathbf o}(\mathbf b)$ are 
        differentiable $(n-1)$-dimensional manifolds, they share the same 
        $(n-1)$-dimensional tangent space at $\mathbf s_{p_e}^{\mathbf o,\mathbf{u}}$. 
        We let $V = \{ \mathbf{v}_{1}, \dots, \mathbf{v}_{n-1} \} \subset \mathbb{R}^{n}$ denote 
        a basis for this tangent space. 

        From the above argument, it is clear that also $\mathbf b$ satisfies both of the criteria in the Lemma, 
        as 1) $\mathbf b \in \Pi_{p_e}(\mathbf{u})$ and 2)
        $(\mathbf s_{p_e}^{\mathbf o,\mathbf u} - \mathbf{b})$ is orthogonal to $\mathcal S^{\mathbf o}_{p_e}(\mathcal{U})$ at $\mathbf s_{p_e}^{\mathbf o,\mathbf u}$
        since $(\mathbf s_{p_e}^{\mathbf o,\mathbf u} - \mathbf{b})$ is orthogonal to $\mathcal{W}^{\mathbf o}(\mathbf b)$ at $\mathbf s_{p_e}^{\mathbf o,\mathbf u}$.
        
        Hence, starting with a pair $(\mathbf a, \mathbf u)$ that satisfies the two conditions of the Lemma, 
        we have identified a point $\mathbf b \in \Pi_{p_e}(\mathbf{u}) \cap \partial \mathcal{B}_{p_e}$ such that $(\mathbf b, \mathbf u)$ satisfies the same conditions. 
        Using that $(\mathbf a, \mathbf u)$ and $(\mathbf b, \mathbf u)$ satisfy these conditions simultaneously, we obtain 
        \begin{center}
            \begin{minipage}{0.55\textwidth}
                \begin{enumerate}    
                    \item $\Rightarrow \mathbf{a}, \mathbf{b} \in \Pi_{p_e}(\mathbf{u}) \Rightarrow \mathbf{a} \cdot \mathbf{u} = \mathbf{b} \cdot \mathbf{u}$,
                    \item $\Rightarrow (\mathbf s_{p_e}^{\mathbf o,\mathbf u} - \mathbf{a}) \cdot \mathbf{v} = (\mathbf s_{p_e}^{\mathbf o,\mathbf u} - \mathbf{b}) \cdot \mathbf{v} = 0$ for any $\mathbf{v} \in V$.
                \end{enumerate}
            \end{minipage}
        \end{center}
        From these conditions we see that $(\mathbf{a} - \mathbf{b}) \cdot \mathbf{u} = 0$ and 
        $(\mathbf{a} - \mathbf{b}) \cdot \mathbf{v} = 0$ for any $\mathbf{v} \in V$.
        Hence, if $\mathbf u$ is linearly independent of $V$, we can conclude that $\mathbf{a} = \mathbf{b}$.

        Assume $\mathbf u = \sum_{i = 1}^{n-1} \alpha_{i} \mathbf{v}_{i}$ for some $\alpha_{1}, \dots, \alpha_{n-1} \in \mathbb{R}$. 
        Then $(\mathbf{b} - \mathbf s_{p_e}^{\mathbf o,\mathbf u}) \cdot \mathbf u = \sum_{i = 1}^{n-1} \alpha_{i} (\mathbf{b} - \mathbf s_{p_e}^{\mathbf o,\mathbf u}) \cdot \mathbf{v}_{i} = 0$.
        Then, by definition of the hyperplane $\Pi_{p_e}(\mathbf u)$, $C^{\mathbf o}_{p_e}(\mathbf u) = (\mathbf{b} - \mathbf{o}) \cdot \mathbf{u} = (\mathbf{b} - \mathbf s_{p_e}^{\mathbf o,\mathbf u} + \mathbf s_{p_e}^{\mathbf o,\mathbf u} - \mathbf{o}) \cdot \mathbf{u} = (\mathbf s_{p_e}^{\mathbf o,\mathbf u} - \mathbf{o}) \cdot \mathbf{u}$.
        But this means that $\mathbf s_{p_e}^{\mathbf o,\mathbf u} \in \Pi_{p_e}(\mathbf u)$, which is impossible.

        We may therefore conclude that $\mathbf{a} = \mathbf{b} \in \Pi_{p_e}(\mathbf{u}) \cap \partial \mathcal{B}_{p_e}$. 
        By the same argument as above, if $\mathbf{b}_{1}$ and $\mathbf{b}_{2}$ are two elements of $\Pi_{p_e}(\mathbf{u}) \cap \partial \mathcal{B}_{p_e}$, 
        then since $(\mathbf{b}_{1}, \mathbf u)$ and $(\mathbf{b}_{2}, \mathbf u)$ both satisfy the 
        conditions of the Lemma, we must have $\mathbf{b}_{1} = \mathbf{b}_{2}$. 
        $\Pi_{p_e}(\mathbf{u}) \cap \partial \mathcal{B}_{p_e}$ is therefore a singleton set, 
        and we can conclude that $\{ \mathbf a \} = \Pi_{p_e}(\mathbf{u}) \cap \partial \mathcal{B}_{p_e}$.

    \qed

    \section{Proof of Proposition \ref{prop:map_to_dB_criterion}}
        \label{app:proof_prop_map_to_dB_criterion}
        We first recall that if $\partial \mathcal{B}_{p_e}$ is a proper convex environmental contour, 
        then for any $\mathbf b \in \partial \mathcal{B}_{p_e}$ there exists some $\mathbf u \in \mathcal{U}$ 
        such that $\mathbf b \in \Pi_{p_e}(\mathbf u)$, and so $\partial \mathcal{B}_{p_e} \subset \cup_{\mathbf u \in \mathcal{U}} \Pi_{p_e}(\mathbf u)$.
        
        Then, if $F : \mathcal{U} \rightarrow \mathbb{R}^{n}$ is a mapping such that the assumptions and conditions 
        of Lemma \ref{lemma:au} hold for any pair $(F(\mathbf u) ,\mathbf u)$, Lemma \ref{lemma:au} lets us conclude that 
        $\{ F(\mathbf u) \} = \Pi_{p_e}(\mathbf{u}) \cap \partial \mathcal{B}_{p_e}$ for any $\mathbf u \in \mathcal{U}$. 

        Hence, $F(\mathcal{U}) = \cup_{\mathbf u \in \mathcal{U}} (\Pi_{p_e}(\mathbf u) \cap \partial \mathcal{B}_{p_e})
        = \partial \mathcal{B}_{p_e} \cap ( \cup_{\mathbf u \in \mathcal{U}} \Pi_{p_e}(\mathbf u)) = \partial \mathcal{B}_{p_e}$.

    \qed

    \section{Proof of Theorem \ref{thm:param}}
        \label{app:proof_thm_param}
        If the the $p_e$-level percentile function $C_{p_{e}}(\textbf u)$ is continuously differentiable 
        on the unit $(n-1)$-sphere, then as 
        $\mathbf s_{p_e}^{\mathbf o} = \mathbf o + 2(C_{p_e}(\mathbf u) - \mathbf u \cdot \mathbf o ) \mathbf u$, 
        the set $\mathcal S^{\mathbf o}_{p_e}(\mathcal{U}) = \{ s_{p_e}^{\mathbf o} \}$ is a differentiable manifold.
        Hence, the assumptions of Lemma \ref{lemma:au} are satisfied. 
        
        We first note that, as a consequence of Lemma \ref{lemma:au}, 
        any supporting hyperplane intersects $\partial \mathcal B_{p_e}$ at a single point, which means that $\mathcal{B}_{p_e}$ is strictly convex.
        For details we refer to the proof of Lemma \ref{lemma:au} in Appendix \ref{app:proof_lemma_au}, 
        where we observe that $\Pi_{p_e}(\mathbf{u}) \cap \partial \mathcal B_{p_e}$ is a singleton set for any $\mathbf u \in \mathcal{U}$, 
        as the pair $(\mathbf b, \mathbf u)$ satisfies the conditions in Lemma \ref{lemma:au} for any $\mathbf{b} \in \Pi_{p_e}(\mathbf{u}) \cap \partial \mathcal B_{p_e}$. 
        (And for any $\mathbf b \in \partial \mathcal B_{p_e}$ we have $\mathbf b \in \Pi_{p_e}(\mathbf{u})$ for some $\mathbf u$ as 
        $\partial \mathcal{B}_{p_e}$ is proper).
        
        We will show that the proposed parametrization in the theorem is valid using Lemma \ref{lemma:au} and Proposition \ref{prop:map_to_dB_criterion}. 
        That is, for any $\mathbf u = \mathbf u(\bm{\theta}) \in \mathcal U$, we must show that
        \begin{center}
            \begin{minipage}{0.55\textwidth}
                \begin{enumerate}    
                    \item $\norm{\mathbf{b}(\bm{\theta}) - \mathbf{o}} = \norm{\mathbf s_{p_e}^{\mathbf o,\mathbf u(\bm{\theta})} - \mathbf{b}(\bm{\theta})}$, and
                    \item $(\mathbf s_{p_e}^{\mathbf o,\mathbf u(\bm{\theta})} - \mathbf{b}(\bm{\theta}))$ is orthogonal to $\mathcal S^{\mathbf o}_{p_e}(\mathcal{U})$ at $\mathbf s_{p_e}^{\mathbf o,\mathbf u(\bm{\theta})}$,
                \end{enumerate}
            \end{minipage}
        \end{center}
        for $\mathbf o \in \mathcal B_{p_e} \setminus \partial \mathcal B_{p_e}$.
        To simplify the notation we will suppress writing out the dependency on $\bm{\theta}$, and write 
        \begin{equation*}
            \mathbf{b} =  C_{p_e}\mathbf{u} + \nabla \mathbf{u} g^{-1} (\nabla C_{p_e})^{T}. 
        \end{equation*}
        Using \eqref{eq:C_conversion} we can express $ \mathbf{b}$ in terms of $ C^{\mathbf o}_{p_e}$:      
        \begin{equation*}
       	\begin{split}
            \mathbf{b} & =  C_{p_e}\mathbf{u} + \nabla \mathbf{u} g^{-1} (\nabla C_{p_e})^{T}\\
            & =  C_{p_e}\mathbf{u} + \mathbf{u}\mathbf{u}^T\mathbf{o} + \nabla \mathbf{u} g^{-1} (\nabla C^{\mathbf o}_{p_e})^{T} + \nabla \mathbf{u} g^{-1} (\nabla \mathbf u )^{T}\mathbf o\\
            & =  \mathbf o + C^{\mathbf o}_{p_e}\mathbf{u} + \nabla \mathbf{u} g^{-1} (\nabla C^{\mathbf o}_{p_e})^{T}, 
           \end{split} 
        \end{equation*}
        where we made use of the property that $ \mathbf{u}\mathbf{u}^T + \nabla \mathbf{u} g^{-1} (\nabla \mathbf u )^{T} = I$ (i.e. the identity operator).
        Note that the metric tensor $g = (\nabla \mathbf{u})^{T}\nabla \mathbf{u}$ is invertible because we have assumed a regular parametrization (and so $\nabla \mathbf{u}$ has full rank). 
        
        To show condition (1) above, we can just compute the norms 
        \begin{equation*}
            \begin{split}
                & \norm{\mathbf b-\mathbf o}^{2} - \norm{\mathbf s_{p_e}^{\mathbf o, \mathbf u} - \mathbf b}^{2} \\
                = &\norm{C^{\mathbf o}_{p_e} \mathbf u + \nabla \mathbf u g^{-1} (\nabla C^{\mathbf o}_{p_e})^{T} }^{2}
                - \norm{C^{\mathbf o}_{p_e} \mathbf u - \nabla \mathbf u g^{-1} (\nabla C^{\mathbf o}_{p_e})^{T}}^{2} \\    
                = & \ 4C^{\mathbf o}_{p_e} \mathbf u \cdot \nabla \mathbf u g^{-1} (\nabla C^{\mathbf o}_{p_e})^{T} \\
                = & \ 0.
            \end{split}
        \end{equation*}
        Here we have used the fact that  $\mathbf u \cdot \nabla \mathbf u = \mathbf u^T\nabla \mathbf u= \frac{1}{2}\nabla( \mathbf u^{T} \mathbf u) = \nabla(1)=0$.
             
        To show condition (2) we will use that the columns of $\nabla \mathbf s^{\mathbf o, \mathbf u}_{p_e}$ 
        form a basis for the tangent space of $\mathcal S^{\mathbf o}_{p_e}(\mathcal{U})$ at $\mathbf s_{p_e}^{\mathbf o,\mathbf u}$.
        The orthogonality condition (2) is therefore equivalent to saying that $\nabla \mathbf (s^{\mathbf o, \mathbf u}_{p_e})^{T} (s^{\mathbf o, \mathbf u}_{p_e} - \mathbf b) = \mathbf 0$.
        But this follows from the definition of $s^{\mathbf o, \mathbf u}_{p_e}$, as
        $\nabla \mathbf s^{\mathbf o, \mathbf u}_{p_e} = \nabla (\mathbf o + 2C^{\mathbf o}_{p_e}\mathbf{u})
        = 2(C^{\mathbf o}_{p_e} \nabla \mathbf u + \mathbf u \nabla C^{\mathbf o}_{p_e})$, and hence
        \begin{equation*}
            \begin{split}
                \frac{1}{2}\nabla \mathbf (s^{\mathbf o, \mathbf u}_{p_e})^{T} (s^{\mathbf o, \mathbf u}_{p_e} - \mathbf b) 
                & = \left( C^{\mathbf o}_{p_e} \nabla \mathbf u^{T} + (\nabla C^{\mathbf o}_{p_e})^{T} \mathbf u^{T} \right)
                \left( C^{\mathbf o}_{p_e} \mathbf u - \nabla \mathbf u A^{-1} (\nabla C^{\mathbf o}_{p_e})^{T} \right) \\ 
                & = (C^{\mathbf o}_{p_e})^{2} \underbrace{\nabla \mathbf u^{T} \mathbf u}_{\mathbf 0} 
                - C^{\mathbf o}_{p_e} \underbrace{ \nabla \mathbf u^{T}  \nabla \mathbf u A^{-1} }_{I} (\nabla C^{\mathbf o}_{p_e})^{T} \\
                & + C^{\mathbf o}_{p_e} (\nabla C^{\mathbf o}_{p_e})^{T} \underbrace{ \mathbf u^{T} \mathbf u }_{1}
                - (\nabla C^{\mathbf o}_{p_e})^{T} \underbrace{ \mathbf u^{T} \nabla \mathbf u }_{\mathbf 0} A^{-1} (\nabla C^{\mathbf o}_{p_e})^{T} \\
                & = - C^{\mathbf o}_{p_e}(\nabla C^{\mathbf o}_{p_e})^{T} + C^{\mathbf o}_{p_e}(\nabla C^{\mathbf o}_{p_e})^{T} \\ 
                & = \mathbf 0.
            \end{split}
        \end{equation*}
        Using Proposition \ref{prop:map_to_dB_criterion} we may then conclude that, given an atlas $\{\mathbf{u}_{i}( \bm{\theta}) \ | \ \bm{\theta} \in \Theta_{i} \}_{i}$ 
        on $\mathcal U$ where each $(\mathbf{u}_{i}, \Theta_{i})$ is a regular parametrization, the corresponding charts 
        $(\mathbf{b}_{i}, \Theta_{i})$ is an atlas on $\partial \mathcal B_{p_e}$. Finally, differentiability of $\partial \mathcal B_{p_e}$ 
        then follows from the given expression for $\mathbf{b}_{i}$ as a function of $\bm{\theta}$. 

    \qed

    \section{Proof of Lemma \ref{lemma:existence}}
    \label{app:proof_lemma_existence}
    
        We first observe that, as a direct consequence of Definition \ref{def:proper_and_valid}, $\mathbf X$ admits a proper convex environmental contour
        if and only if every hyperplane $\Pi_{p_e}(\mathbf u)$ is a supporting hyperplane of $\mathcal{B}_{p_e}$. 
        That is, if and only if $\mathcal{B}_{p_e} \cap \Pi_{p_e}(\mathbf u) \neq \emptyset$ for all $\mathbf u \in \mathcal{U}$.
        
        Hence, if $\mathbf X$ admits a proper convex environmental contour, we can select 
        $\mathbf o \in \Pi_{p_e}(\mathbf u') \cap \partial \mathcal B_{p_e}$ which (by Lemma \ref{lemma:cont_vor}) satisfies the condition.
        
        If $\mathbf X$ does not admit a proper convex environmental contour, then there is 
        some hyperplane $\Pi_{p_e}(\mathbf u')$ that does not intersect $B_{p_e}$. Hence, for any 
        $\mathbf o \in \Pi_{p_e}(\mathbf u')$ we have $\mathbf o \notin B_{p_e}$, and by Lemma \ref{lemma:cont_vor}
        there must exist some $\mathbf u^*$ where $C^{\mathbf o}_{p_e}(\mathbf u^*) < 0$.
    
    \qed

    \section{Proof of Lemma \ref{lemma:b_theta_pi}}
    \label{app:proof_lemma_b_theta_pi}

        Dropping the dependency on $\bm \theta$ and $p_e$ for simpler notation, we may write
        \begin{equation*}
            \mathbf u^T \mathbf b = \mathbf u^T  (C \mathbf u + \nabla \mathbf u g^{-1} \nabla C^T) 
            = C \mathbf u^T \mathbf u + \mathbf u^T \nabla \mathbf u g^{-1} \nabla C^T = C,
        \end{equation*}
        as $\mathbf u^T \mathbf u = 1$ and $\mathbf u^T\nabla \mathbf u= \frac{1}{2}\nabla( \mathbf u^{T} \mathbf u) = \nabla(1)=0$. 
        This means that $\mathbf b (\bm{\theta}) \in \Pi(\bm{\theta})$.
        Similarly, we observe that 
        \begin{equation*}
            \nabla \mathbf u^T \mathbf b = 
            C \nabla \mathbf u^T \mathbf u + \nabla \mathbf u^T \nabla \mathbf u g^{-1} \nabla C^T = \nabla C^T,
        \end{equation*}
        as $\nabla \mathbf u^T \nabla \mathbf u = g$ by definition. 
        From the chain rule we then get 
        $\mathbf u^T \nabla \mathbf b = \nabla (\mathbf u^T \mathbf b) - (\nabla \mathbf u^T \mathbf b)^T
        = \nabla C - \nabla C = \bm{0}$. Since the hyperplane $\Pi(\bm{\theta})$ has normal vector $\mathbf{u}(\bm{\theta})$, we can conclude that $\Pi(\bm{\theta})$ is tangential to 
        $\mathbf b (\Theta)$ at $\mathbf{b}(\bm{\theta})$.

    \qed 
    
    \section{Proof of Theorem \ref{thm:existence}}
    \label{app:proof_thm_existence}

        To simplify notation, we drop the dependency $p_e$ and the index $i$ of the parametrization.
        
        Assume $(2)$ is true and let $\mathcal{B}$ denote the closed convex set. Then Lemma \ref{lemma:b_theta_pi} implies that all hyperplanes $\Pi(\bm{\theta})$ 
        are supporting hyperplanes of $\mathcal{B}$, and so $\partial \mathcal{B}$ is a 
        proper convex environmental contour. The fact that $(1) \Rightarrow (2)$ comes as a direct consequence of Theorem \ref{thm:param}, so we have that $(1) \Leftrightarrow (2)$.
    
        To show that $(1) \Rightarrow (3)$, we first note that when $\mathbf X$ admits a proper convex environmental contour, then since $\bm{b} (\bm{\theta}) \in \Pi(\bm{\theta})$ (see Lemma \ref{lemma:b_theta_pi}) it follows from Lemma \ref{lemma:cont_vor} that 
        $\kappa(\bm{\theta} | \bm{\theta}') \geq 0$ for all $\bm{\theta}$ and $\bm{\theta}'$.
        For the converse, assume that $\mathbf X$ does not admit a proper convex environmental contour. 
        Then from Lemma \ref{lemma:existence} there exists some $\mathbf u'$ such that 
        for any $\mathbf o \in \Pi(\mathbf u')$ we can find some $\mathbf u$
        where $C^{\mathbf o} (\mathbf u) < 0$.
        In forms of the given parametrization, this means that we can find some $\bm{\theta}$ and $\bm{\theta}'$ 
        where $C^{\mathbf o} (\bm{\theta}) < 0$ for any $\mathbf o \in \Pi(\bm{\theta}')$.
        As $\mathbf{b}(\bm{\theta}') \in \Pi(\bm{\theta}')$ 
        we have that $\kappa(\bm{\theta} | \bm{\theta}') = C^{\mathbf{b}(\bm{\theta}')}(\bm{\theta}) <0$. Hence $(1) \Leftrightarrow (3)$.
        
        Finally, $(3) \Leftrightarrow (4)$ follows from the fact that $\mathbf{b}(\bm{\theta}') \in \Pi(\bm{\theta}')$ which means that $\kappa(\bm{\theta}' | \bm{\theta}') = 0$.
    
    \qed

    \section{Proof of Corollary \ref{cor:hess}}
    \label{app:proof_cor_hess}
        
        From statement $(4)$ in Theorem \ref{thm:existence}, $\kappa(\bm{\theta} | \bm{\theta}')$ attains a local minimum at $\bm{\theta} = \bm{\theta}'$, which means that 
        the matrix $A(\bm \theta) = \nabla_{\bm \theta}\nabla_{\bm \theta} \kappa(\bm{\theta} | \bm{\theta}') |_{\bm{\theta} = \bm{\theta}'}$ is
        positive semi-definite $\forall \bm \theta \in \Theta$. 
        Suppressing the notation $\bm \theta$ and $p_e$ we can write 
        \begin{equation}
        \label{eq:A1}
        \begin{split}
            A & = \nabla\nabla C + \mathbf b^T \nabla\nabla \mathbf u \\
            & =  \nabla\nabla C + \left(  C\mathbf{u} + \nabla \mathbf{u}g^{-1}\nabla C^{T} \right)^T \nabla\nabla \mathbf u\\
            & =  \nabla\nabla C + C\mathbf{u}^T  \nabla\nabla \mathbf u
            + (\nabla \mathbf{u} g^{-1}\nabla C^{T})^T \nabla\nabla \mathbf u.
        \end{split}
        \end{equation}
        The second term in the last line of \eqref{eq:A1} above can be rewritten in terms of the metric tensor $g$:
        \begin{equation*}
            \begin{split}
                C\mathbf{u}^T \nabla\nabla \mathbf u 
                = C \nabla \left( \mathbf{u}^T \nabla \mathbf u\right) -  C(\nabla \mathbf{u})^T \nabla \mathbf u 
                = -g C,
            \end{split}
        \end{equation*}
        because $\mathbf{u}^T \nabla \mathbf u = \bm 0$ and $ (\nabla \mathbf{u})^T \nabla \mathbf u = g$.
    
        The third term in the last line of \eqref{eq:A1} can be expressed as $\nabla C g^{-T} (\nabla \mathbf{u})^T  \nabla\nabla \mathbf u$.
        In index form (using Einstein summation convention) we may write the matrix elements of this term as $c_{,m}g^{lm}u_{k,l} u_{k,ij} = c_{,m}g^{lm}\Gamma_{lij} = c_{,m}\Gamma^m_{ij}$, where we have recognised the Christoffel symbols of the first and second kind, i.e. $\Gamma_{lij} =u_{k,l} u_{k,ij}$ and $\Gamma^m_{ij}=g^{lm}\Gamma_{lij}$.
        Therefore we may write 
        \begin{equation}
            A_{ij}(\bm \theta) = \left(  \frac{\partial C(\bm \theta)}{\partial \theta_i \partial \theta_j} -\Gamma^m_{ij}\frac{C(\bm \theta)}{\partial \theta_m}  \right) + g_{ij}(\bm \theta) C(\bm \theta).
        \end{equation}
        The term in brackets correspond to the Hessian on a Riemann manifold, and we may therefor write 
        \begin{equation}
            A(\bm \theta) = Hess( C(\bm \theta)) + g(\bm \theta)  C(\bm \theta) .
        \end{equation}

    \qed 
    
\end{appendices}

\bibliography{references,Referanseliste_Erik}

\begin{thebibliography}{10}
\expandafter\ifx\csname url\endcsname\relax
  \def\url#1{\texttt{#1}}\fi
\expandafter\ifx\csname urlprefix\endcsname\relax\def\urlprefix{URL }\fi
\expandafter\ifx\csname href\endcsname\relax
  \def\href#1#2{#2} \def\path#1{#1}\fi

\bibitem{DNVGLRP-C205}
D.~GL, Environmental Conditions and Environmental Loads, DNV GL, september 2019
  Edition, {DNVGL-RP-C205} (2019).

\bibitem{NORSOK_N003_17}
NORSOK, {NORSOK Standard N-003:2017. Action and action effects}, edition 3
  (2017).

\bibitem{Haver80}
S.~Haver, Analysis of uncertainties related to the stochastic modelling of
  ocean waves, Tech. Rep. UR-80-09, Norges tekniske h{\o}gskole (1980).

\bibitem{Haver87}
S.~Haver, On the joint distribution of heights and periods of sea waves, Ocean
  Engineering 14 (1987) 359--376.

\bibitem{WitUCBH93}
S.~Winterstein, T.~Ude, C.~Cornell, P.~Bjerager, S.~Haver, Environmental
  parameters for extreme response: Inverse {FORM} with omission factors, in:
  Proc. 6th International Conference on Structural Safety and Reliability,
  1993.

\bibitem{HW:ENvContLin09}
S.~Haver, S.~Winterstein, Environmental contour lines: A method for estimating
  long term extremes by a short term analysis, Transactions of the Society of
  Naval Architects and Marine Engineers 116 (2009) 116--127.

\bibitem{Leira:StocProcCont08}
B.~J. Leira, A comparison of stochastic process models for definition of design
  contours, Structural Safety 30 (2008) 493--505.

\bibitem{NdLY:EstExtrRespEnvCont98}
J.~M. Niedzwwecki, J.~van~de Lindt, J.~Yao, Estimating extreme tendon response
  using environmental contours, Engineering Structures 20 (1998) 601--607.

\bibitem{WJK:RelFloatStructLoadFactDesign99}
S.~R. Winterstein, A.~K. Jha, S.~Kumar, Reliability of floating structures:
  Extreme response and load factor design, Journal of Waterway, Port, Coastal
  and Ocean Engineering 125 (1999) 163--169.

\bibitem{BM:ApplyContHullLoads01}
G.~S. Baarholm, T.~Moan, Application of contour line method to estimate extreme
  ship hull loads considering operational restrictions, Journal of Ship
  Research 45 (2001) 228--240.

\bibitem{SM:DesignLoadWindTurbEnvCont06}
K.~Saranyasoontorn, L.~Manuel, Design loads for wind turbines using the
  environmental contour method, in: 44th AIAA Aerospace Sciences Meeting and
  Exhibit, American Institute of Aeronautics and Astronautics (AIAA), 2006, pp.
  AIAA 2006--1365.

\bibitem{BHL:OMAE2007-29417}
G.~S. Baarholm, H.~Sverre, C.~M. Larsen, Wave sector dependent contour lines,
  in: Proc. 26th International Conference on Offshore Mechanics and Arctic
  Engineering (OMAE 2007), American Society of Mechanical Engineers (ASME),
  2007.

\bibitem{BH_EnvContSum09}
G.~S. Baarholm, S.~Haver, Application of environmental contour lines - a
  summary of a number of case studies, in: Proc. International Conference on
  Floating Structures for Deepwater Operations, ASRANet, 2009.

\bibitem{bhoe:CombContHsT10}
G.~S. Baarholm, S.~Haver, O.~D. {\O}kland, Combining contours of significant
  wave height and peak period with platform response distributions for
  predicting design response, Marine Structures 23 (2010) 147--163.

\bibitem{OMAE2011-49886}
P.~Jonathan, K.~Ewans, J.~Flynn, On the estimation of ocean engineering design
  contours, in: Proc. 30th International Conference on Ocean, Offshore and
  Arctic Engineering (OMAE 2011), American Society of Mechanical Engineers
  (ASME), 2011.

\bibitem{HaverWAveWorkshop13}
S.~Haver, K.~Bruserud, Environmental contour method: An approximate method for
  obtaining characteristic response extremes for design purposes, in: Proc.
  13th International Workshop on Wave Hindcasting and Forecasting \& 4th
  Coastal Hazard Symposium, 2013.

\bibitem{MGM:ApplContTwoBodFloatEngConv13}
M.~J. Muliawan, Z.~Gao, T.~Moan, Application of the contour line method for
  estimating extreme responses in the mooring lines of a two-body floating wave
  energy converter, Journal of Offshore Mechanics and Arctic Engineering 135
  (2013) 031301:1--10.

\bibitem{OMAE2015-41680}
C.~Armstrong, C.~Chin, I.~Penesis, Y.~Drobyshevski, Sensitivity of vessel
  response to environmental contours of extreme sea states, in: Proc. 34th
  International Conference on Ocean, Offshore and Arctic Engineering (OMAE
  2015), American Society of Mechanical Engineers (ASME), 2015.

\bibitem{LW:DynamicIFORM16}
L.~D. Lutes, S.~R. Winterstein, A dynamic inverse {FORM} method: Design
  contours for load combination problems, Probabilistic Engineering Mechanics
  44 (2016) 118--127.

\bibitem{LGM:ModEnvCont16}
Q.~Li, Z.~Gao, T.~Moan, Modified environmental contour method for predicting
  long-term extreme responses of bottom-fixed offshore wind turbines, Marine
  Structures 48 (2016) 15--32.

\bibitem{E-GSDN:PCAContours16}
A.~C. Eckert-Gallup, C.~J. Sallaberry, A.~R. Dallman, V.~S. Neary, Application
  of principal component analysis ({PCA}) and imrpoved joint probability
  distributions to the inverse first-order reliability method ({I-FORM}) for
  predicting extreme sea states, Ocean Engineering 112 (2016) 307--319.

\bibitem{VanemSeasonalContour17}
E.~Vanem, A simple approach to account for seasonality in the description of
  extreme ocean environments, Marine Systems \& Ocean Technology 13 (2018)
  63--73.

\bibitem{HusebyVB:ESREL2019}
A.~B. Huseby, E.~Vanem, M.~H. Barbosa, Environmental contours for mixtures of
  distributions, in: Proc. ESREL 2019, European Safety and Reliability
  Association(ESRA), 2019.

\bibitem{HK:EnvContCircLin17}
Z.~S. Haghayeghi, M.~J. Ketabdari, Development of environmental contours for
  circular and linear metocean variables, International Journal of Renewable
  Energy Research 7 (2017) 682--693.

\bibitem{S-GH-ZM-I:EnvContNATAF13}
F.~Silva-Gonz\'{a}lez, E.~Heredia-Zavoni, R.~Montes-Iturrizaga, Development of
  environmental contours using {N}ataf distribution model, Ocean Engineering 58
  (2013) 27--34.

\bibitem{M-IH-Z:EnvContCop15}
R.~Montes-Iturrizaga, E.~Heredia-Zavoni, Environmental contours using copulas,
  Applied Ocean Research 52 (2015) 125--139.

\bibitem{DH:NewEnvCont:OMAE2019-95993}
Q.~Derbanne, G.~da~Hauteclocque, A new approach for environmental contour and
  multivariate de-clustering, in: Proc. 38th International Conference on Ocean,
  Offshore and Arctic Engineering (OMAE 2019), American Society of Mechanical
  Engineers (ASME), 2019.

\bibitem{DahlHuseby:ESREL18}
K.~R. Dahl, A.~B. Huseby, Buffered environmental contours, in: Proc. ESREL
  2018, European Safety and Reliability Association(ESRA), 2018.

\bibitem{CL:ISORMContours18}
W.~Chai, B.~J. Leira, Environmental contours based on inverse {SORM}, Marine
  Structures 60 (2018) 34--51.

\bibitem{HOWT:HighDensityContour2017}
A.~F. Haselsteiner, J.-H. Ohlendorf, W.~Wosniok, K.-D. Thoben, Deriving
  environmental contours from highest density regions, Coastal Engineering 123
  (2017) 42--51.

\bibitem{M-IHZ:UncertEnvCont17}
R.~Montes-Iturrizaga, E.~Heredia-Zavoni, Assessment of uncertainty in
  environmental contours due to parametric uncertainty in models of the
  dependence structure between metocean variables, Applied Ocean Research 64
  (2017) 86--104.

\bibitem{VanemGB-G:ContourUncert18}
E.~Vanem, O.~Gramstad, E.~M. Bitner-Gregersen, A simulation study on the
  uncertainty of environmental contours due to sampling variability for
  different estimation methods, Applied Ocean Research 91 (2019) 101870.

\bibitem{Vanem:CombinedContours19}
E.~Vanem, Environmental contours for describing extreme ocean wave conditions
  based on combined datasets, Stochastic Environmental Research and Risk
  Assessment 33 (2019) 957--971.

\bibitem{MNCCE-GM:AltApproachEnvCont18}
L.~Manuel, P.~T. Nguyen, J.~Canning, R.~G. Coe, A.~C. Eckert-Gallup, N.~Martin,
  Alternative approaches to develop environmental contours from metocean data,
  Journal of Ocean Engineering and Marine Energy 4 (2018) 293--310.

\bibitem{ECSADESjointPaper19}
E.~Ross, O.~C. Astrup, E.~Bitner-Gregersen, N.~Bunn, G.~Feld, B.~Gouldby,
  A.~Huseby, Y.~Liu, D.~Randell, E.~Vanem, P.~Jonathan, On environmental
  contours for marine and coastal design, Ocean Engineering 195 (2020) 106194.

\bibitem{Vanem:EnvCont12}
A.~B. Huseby, E.~Vanem, B.~Natvig, A new approach to environmental contours for
  ocean engineering applications based on direct {M}onte {C}arlo simulations,
  Ocean Engineering 60 (2013) 124--135.

\bibitem{Vanem:EnvCont14}
A.~B. Huseby, E.~Vanem, B.~Natvig, Alternative environmental contours for
  structural reliability analysis, Structural Safety 54 (2015) 32--45.

\bibitem{HusebyVN:ESREL2014}
A.~B. Huseby, E.~Vanem, B.~Natvig, A new {M}onte {C}arlo method for
  environmental contour estimation, in: Proc. ESREL 2014, European Safety and
  Reliability Association(ESRA), 2014.

\bibitem{HusebyVE:ESREL2017}
A.~B. Huseby, E.~Vanem, K.~Eskeland, Evaluating properties of environmental
  contours, in: Proc. ESREL 2017, European Safety and Reliability
  Association(ESRA), 2017.

\bibitem{VanemB-G:JOMAE15}
E.~Vanem, E.~M. Bitner-Gregersen, Alternative environmental contours for marine
  structural design - a comparison study, Journal of Offshore Mechanics and
  Arctic Engineering 137 (2015) 051601:1--8.

\bibitem{VanemContourCompStruc17}
E.~Vanem, A comparison study on the estimation of extreme structural response
  from different environmental contour methods, Marine Structures 56 (2017)
  137--162.

\bibitem{ECSADESShipCase19}
E.~Vanem, B.~Guo, E.~Ross, P.~Jonathan, Comparing different contour methods
  with response-based methods for extreme ship response analysis, Marine
  Structures 69 (2919) 102680.

\bibitem{NFPQ-R:JointDistMultVarIFORMCont07}
R.~Nerzic, C.~Frelin, M.~Prevesto, V.~Quiniou-Ramus, Joint distribution of
  wind/waves/current in {W}est {A}frica and derivation of multivariate extreme
  {I-FORM} contours, in: Proc. 17th International Offshore and Polar
  Engineering Conference (ISOPE 2007), The International Society of Offshore
  and Polar Engineering (ISOPE), 2007.

\bibitem{dLN:EnvContEarthquake00}
J.~van~de Lindt, J.~Niedzwecki, Environmental contour analysis in earthquake
  engineering, Engineering Structures 22 (2000) 1661--1676.

\bibitem{SM:ModWindTurbineExtrLoad04}
K.~Saranyasoontorn, L.~Manuel, Efficient models for wind turbine extreme loads
  using inverse reliability, Journal of Wind Engineering and Industrial
  Aerodynamics 92 (2004) 789--804.

\bibitem{OFQ:RelRespFPSOEnvCont07}
P.~Orsero, E.~Fontaine, V.~Quiniou, Reliability and response based design of a
  moored {FPSO} in {W}est {A}frica using multivariate environmental contours
  and response surfaces, in: Proc. 17th International Offshore and Polar
  Engineering Conference (ISOPE 2007), The International Society of Offshore
  and Polar Engineering (ISOPE), 2007.

\bibitem{M-IH-Z:MultivarEnvContCvineCop16}
R.~Montes-Iturrizaga, E.~Heredia-Zavoni, Multivariate environmental contours
  using {C}-vine copulas, Ocean Engineering 118 (2016) 68--82.

\bibitem{Vanem3Dcontour17}
E.~Vanem, 3-dimensional environmental contours based on a direct sampling
  method for structural reliability analysis of ships and offshore structures,
  Ships and Offshore Structures 14 (2018) 74--85.

\bibitem{RMP:3DExtrValModTension19}
N.~Raillard, M.~Prevesto, H.~Pineau, 3-d environmental extreme value models for
  the tension in a mooring line of a semi-submersible, Ocean Engineering 184
  (2019) 23--31.

\bibitem{Leonard:2015:Geometry_of_convex_sets}
I.~Leonard, J.~Lewis, Geometry of Convex Sets, Wiley, 2015.

\bibitem{Okabe:2000:Spatial_Tesselations}
A.~Okabe, B.~Boots, K.~Sugihara, S.~N. Chiu, Spatial Tessellations: Concepts
  and Applications of {V}oronoi Diagrams, 2nd Edition, Series in Probability
  and Statistics, John Wiley and Sons, Inc., 2000.

\bibitem{Schaller:2013:Set_Voronoi}
F.~M.~Schaller, S.~Kapfer, M.~Evans, M.~J.F.~Hoffmann, T.~Aste, M.~Saadatfar,
  K.~Mecke, G.~W.~Delaney, G.~Schröder-Turk, Set {V}oronoi diagrams of 3{D}
  assemblies of aspherical particles, Philosophical Magazine 93.

\bibitem{Aurenhammer:1991:VDS}
F.~Aurenhammer, Voronoi diagrams - a survey of a fundamental geometric data
  structure, ACM Comput. Surv. 23~(3) (1991) 345--405.

\bibitem{Huseby:stk4400}
A.~B. Huseby, K.~R. Dahl,
  \href{https://www.uio.no/studier/emner/matnat/math/STK4400/h18/notater/week-41.pdf}{{Lecture
  notes STK4400 - Risk and reliability analysis}} (09 2018).
\newline\urlprefix\url{https://www.uio.no/studier/emner/matnat/math/STK4400/h18/notater/week-41.pdf}

\bibitem{Marsaglia:1972:nsphere}
G.~Marsaglia, Choosing a point from the surface of a sphere, Ann. Math.
  Statist. 43~(2) (1972) 645--646.

\bibitem{ElzbietaBook12}
E.~Bitner-Gregersen, Joint long term models of met-ocean parameters, in:
  C.~Guedes~Soares (Ed.), Marine Technology and Engineering: CENTEC Anniversary
  Book, CRC Press, 2012.

\bibitem{EBG:JointDesc2015}
E.~M. Bitner-Gregersen, Joint met-ocean description for design and operation of
  marine structures, Applied Ocean Research 51 (2015) 279--292.

\bibitem{BGH:JointEnvModRel91}
E.~Bitner-Gregersen, S.~Haver, Joint environmental model for reliability
  calculations, in: Proc. 1st International Offshore and Polar Engineering
  conference (ISOPE 1991), The International Society of Offshore and Polar
  Engineering (ISOPE), 1991.

\bibitem{BGH:JointEnvParam89}
E.~Bitner-Gregersen, S.~Haver, Joint long term description of environmental
  parameters for structural response calculation, in: Proc. 2nd International
  Workshop on Wave Hindcasting and Forecasting, 1989.

\bibitem{Shephard:1964shadow}
G.~C. Shephard, Shadow systems of convex sets, Israel Journal of Mathematics
  2~(4) (1964) 229--236.

\bibitem{Epstein:shadow}
C.~l.~Epstein, \href{https://www.math.upenn.edu/~cle/papers/slatgm.pdf}{{Convex
  regions, shadows, and the Gauss map}}.
\newline\urlprefix\url{https://www.math.upenn.edu/~cle/papers/slatgm.pdf}

\bibitem{Martini:2019bodies}
H.~Martini, L.~Montejano, D.~Oliveros, Bodies of constant width, Springer,
  2019.

\end{thebibliography}

\end{document}